\documentclass[compsoc,10pt]{IEEEtran}
\usepackage[labelfont=bf,textfont={bf}]{caption}

\usepackage{dblfloatfix}
\usepackage{tabularx}
\usepackage{booktabs}
\usepackage[numbers]{natbib}
\usepackage{amsmath}
\usepackage{url}
\usepackage{colortbl}
\usepackage[table]{xcolor}
\usepackage{rotating}
\usepackage{amssymb}
\usepackage{blindtext}
\usepackage{wrapfig}
\usepackage{graphicx}
\usepackage{csvsimple}
\usepackage{algorithm}
\usepackage[noend]{algpseudocode}
\usepackage{float}
\usepackage{aliascnt}
\usepackage{multirow}
\usepackage{soul}
\usepackage{color, colortbl}
\usepackage{babel,adjustbox,booktabs,multirow}
\usepackage[most]{tcolorbox}
\usepackage{subcaption}
\usepackage{balance}
\usepackage{pifont}
\usepackage{changepage}
\usepackage{framed}

\newcommand{\tick}{\ding{51}}
\newcolumntype{L}{>{\raggedright\arraybackslash}X}

\algnewcommand\algorithmicforeach{\textbf{foreach}}
\algdef{S}[FOR]{ForEach}[1]{\algorithmicforeach\ #1\ \algorithmicdo}

\newaliascnt{eqfloat}{equation}
\newfloat{eqfloat}{h}{eqflts}
\floatname{eqfloat}{Equation}

\newtcolorbox{myquote}{colback=lightgray!20!white,colframe=yellow!75!black,grow to right by=-10mm,grow to left by=-10mm,
    boxrule=0pt,boxsep=0pt,breakable} \makeatletter

\newcommand*{\ORGeqfloat}{}
\let\ORGeqfloat\eqfloat
\def\eqfloat{%
  \let\ORIGINALcaption\caption
  \def\caption{%
    \addtocounter{equation}{-1}%
    \ORIGINALcaption
  }%
  \ORGeqfloat
}

\makeatletter
 {\par\unskip\endMakeFramed}
\makeatother

\definecolor{formalshade}{rgb}{0.93,0.93,0.93}

\definecolor{darkblue}{rgb}{0.2, 0.2, 0.2}

\newenvironment{formal}{%
  \def\FrameCommand{%
    \hspace{1pt}%
    {\color{darkblue}\vrule width 2pt}%
    {\color{formalshade}\vrule width 4pt}%
    \colorbox{formalshade}%
  }%
  \MakeFramed{\advance\hsize-\width\FrameRestore}%
  \noindent\hspace{-1pt}
  \begin{adjustwidth}{}{7pt}%
  \vspace{2pt}\vspace{2pt}%
}
{%
  \vspace{3pt}\end{adjustwidth}\endMakeFramed%
}

\newcommand{\rqn}[1]{\underline{\textbf{RQ#1:}}}

\usepackage{enumitem}
\setlist{nosep} 
\newcommand{\bi}{\begin{itemize}}
	\newcommand{\ei}{\end{itemize}}
\newcommand{\be}{\noindent\begin{enumerate}}
	\newcommand{\ee}{\noindent\end{enumerate}}

\usepackage[tikz]{bclogo}
{\noindent\begin{minipage}[c]{\linewidth}%
\begin{bclogo}[couleur=gray!20,%
                arrondi=0,logo=\none,%
                ombre=true%
                ]{{\small  ~#1}}}%
{\end{bclogo}\vspace{2mm}\end{minipage}}

\newcommand{\respto}[1]{}

\newcommand{\BLUE}{\color{black}}
\newcommand{\BLACK}{\color{black}}
\newcommand{\ORANGE}{\color{orange}}

\begin{document}


\title{How to Find Actionable Static Analysis Warnings: A Case Study with FindBugs}

\author{Rahul Yedida, 
Hong Jin Kang,
Huy Tu, Xueqi Yang, David Lo~\IEEEmembership{Fellow,~IEEE}, 
        Tim~Menzies,~\IEEEmembership{Fellow,~IEEE}
\IEEEcompsocitemizethanks{\IEEEcompsocthanksitem R. Yedida, X. Yang and  T. Menzies are with the Department
of Computer Science, North Carolina State University, Raleigh, USA.
 \protect\\
E-mail: ryedida@ncsu.edu, xyang37@ncsu.edu, timm@ieee.org
\IEEEcompsocthanksitem H.J. Kang and D. Lo are with the School of Computing and Information Systems, Singapore Management University, Singapore.
\protect\\
E-mail: hjkang.2018@smu.edu.sg, davidlo@smu.edu.sg
\IEEEcompsocthanksitem  H. Tu is with Meta Platforms, Inc. 
E-mail: huyqtu7@gmail.com}}

\markboth{IEEE Transactions on Software Engineering}%
{Yedida \MakeLowercase{\textit{et al.}}: How to Recognize and Avoid Static Code Analysis False Alarms}

\IEEEtitleabstractindextext{
\begin{abstract}
Automatically generated static
code warnings  suffer from
a large number of false alarms.
Hence, developers
only take action
on a small percent of those warnings. 
To better predict which  static code warnings
should {\em not} be ignored, 
  we suggest that      analysts
  need to look deeper
  into their algorithms to find choices
 that better improve the particulars of their specific problem. Specifically,
 we show here that effective predictors
 of such warnings can be created   by methods
that {\em locally adjust} the decision boundary
(between actionable warnings and others).
 These  methods yield a  new high water-mark for recognizing actionable static code warnings.
For eight open-source Java projects
(cassandra, jmeter, commons, lucene-solr, maven,
ant, tomcat, derby)
 we    achieve perfect test results on 4/8 datasets
 and, overall, a median AUC (area under the true negatives, true positives curve) of 92\%.

\end{abstract}

}

\maketitle

%
\IEEEpeerreviewmaketitle


\section{Introduction}\label{intro}

Static analysis (SA) tools report errors
in source code, without needing to execute that code. This makes them very popular in industry. For example, the FindBugs tool~\cite{ayewah2010google}
has been downloaded over a million times. Unfortunately, due to the imprecision of static analysis and the different contexts where bugs appear, SA tools often suffer from
a large number of false alarms that are deemed to be not actionable~\cite{tomassi2021real}.
Hence, developers never act on most of 
their warnings \cite{heckman2008establishing, heckman2009model, kim2007warnings}. 
 Previous research work shows that   35\% to
91\% of SA warnings reported as bugs by SA tools 
are routinely
ignored by developers~\cite{heckman2009model,heckman2008establishing,kim07}.

Those  false alarms produced by SA tools are a significant barrier to the wide-scale adoption of these SA tools ~\cite{johnson2013don,ChristakisB16}. 
Accordingly, in 2018~\cite{wang2018there}, 2020~\cite{yang2021learning}
and 2021~\cite{yang2021understanding}, Wang et al. and Yang et al. proposed
data miners that  found the subset of static code warnings that
developers found ``actionable'' (i.e. those that
motivate developers to change the code). 
But in 2022, Kang et al.~\cite{kang2022detecting} showed that of the
31,000+ records used by  
Wang et al. and Yang et al., they could only generate 768  error-free records-- which meant all the prior Wang and Yang et al. results need to be revisited.

When Kang et al. tried to build predictors from the 768 good records, they found that their   best-performing predictors were not effective
(e.g., very low median AUCs of 41\%), for details see 
Table~\ref{tab:initial_svm}. 
  Hence the following remains an open research question:

\begin{formal}\noindent\rqn{1} \textit{For detecting actionable static code warnings,
what data mining methods should we recommend?}
\end{formal}
\noindent
This paper conjectures that prior work failed to find good predictors because of a {\em locality problem}. In the learners
used in that prior work, the decision boundary between  
 actionable warnings and other
was determined by a single {\em global policy}. 
 In detail, changes to the values of the hyper-parameters of   (e.g.) an SVM learner (used by Kang et al.~\cite{kang2022detecting})
make \textit{global} changes to the decision boundary (i.e., the global shape of the decision boundary is modified); instead, we argue for \textit{local} changes to the decision boundary. This allows us to make different local adjustments at different regions of the decision boundary to adapt it to the local data.
 
More specifically, we conjecture  that:
 \begin{quote}
 {\em 
 For complex data, \underline{\bf global} treatments  
   perform worse  than \underline{\bf localized} treatments
 which   adjust different parts of the  landscape in 
 different ways.}
 \end{quote}
 To test this, we use    {\em local} treatments to adjust the decision boundary
  in different ways in different parts of the data.
 \be
\item
{\em Boundary engineering}:  adjust the decision boundary near our data points;
\item
{\em Label engineering}:   control outliers
in a local region by using just a small fraction of those local labels;
\item
{\em Instance engineering}: addressing class imbalance in local regions of the data;
\item
These treatments are combined with
{\em parameter engineering}  to control   how we   build models.
 \ee
We call this combination of   treatments
GHOST2 (GHOST2 extends GHOST~\cite{yedida2021value} which just used one of these treatments).  When researchers propose an intricate
combination of 
ideas, it is prudent to ask several questions:
   
\begin{formal}\noindent 
\rqn{2} {\em Does GHOST2's  combination of {\em instance}, {\em label}, {\em boundary}  and {\em parameter} engineering, reduce the complexity of  the decision boundary?}
\end{formal}

Later in this paper, we will show evidence that our proposed methods
simplifies the ``error landscape'' of a data set (a concept which we will discuss, in detail in \S\ref{rx}).


\begin{formal}\noindent 
\rqn{3} {\em Does  GHOST2's  use of {\em instance}, {\em label}, {\em boundary}  and {\em parameter} treatments improve predictions?} 
\end{formal}
  
 Using data from Kang et al. (768 records from  eight open-source Java projects), we show that
 GHOST2  was able to generate excellent predictors for actionable static code warnings.

 \begin{formal}\noindent   
\rqn{4} {\em Are all parts of GHOST2 necessary; i.e. would something
simpler also achieve the overall goal?}
\end{formal}

   To answer {\bf RQ4},   this paper   reports an 
{\em ablation study} that removes
one treatment at a time from our four recommended treatments. For the purposes of 
recognizing and avoiding
static code analysis false alarms, it will be shown that,
ignoring any part of our proposed solution   leads
to worse predictors.
Hence, while we do not know if  changes to our design might lead to {\em better}
predictors, the
ablations study does show that removing anything from that design makes matters {\em worse}.

This work has six key contributions:
\be
\item
As a way to address, in part, the methodological problems raised by Kang et al. GHOST2 makes its conclusions using   a small percentage of the raw data (10\%). That is, to address the issues
of corrupt data found by Kang et al., we say ``use less data'' and, for the data that is used,   ``reflect more on that data''.
\item
A case study of successful open collaboration
by software analytics researchers. 
This paper is joint work between the Yang et al. and Kang et al. teams
from the United States and Singapore. By recognizing a shared problem, then sharing  data and tools, in eight weeks these two teams produced a new
state-of-the-art result that improves on all of the past  
papers by these two teams (within this research arena). This illustrates the value of open and collaborative science, where groups with different initial findings come together to help each other in improving the state-of-the-art for the benefit of science and industry.
\item
Motivation for changing the way we train  software analytics newcomers.  It is insufficient to just reflect on the
different properties of off-the-shelf learners.
Analysts may need to be skilled in
boundary, label, parameter and instance engineering.
\item GHOST2's  design, implementation, and evaluation.
\item A new high-water mark in software analytics for learning actionable static code warnings.
\item A reproduction package  that other researchers can use to repeat/refute/improve on our results\footnote{ \url{https://github.com/yrahul3910/static-code-warnings/}}.

\ee
The rest of this paper is structured as follows. The next section offers some background notes. \S \ref{problem} discusses the locality
problem for complex data sets and \S \ref{rx}
offers details on our treatments.  \S \ref{sec:methods} describes our experimental methods
after which, in \S \ref{sec:results}, we show that GHOST2      outperforms (by a large margin) prior results from Kang et al. We discuss threats to validity for our study in \S \ref{sec:threats}, before a discussion in \S \ref{sec:discussion} and concluding in \S \ref{sec:conclusion}.

Before all that,  we digress to   make two important points. 
\bi
\item
A  learned model must be tested on the kinds of data expected in practice. 
Hence,  
  any treatments to the data  (e.g.  via instance, label, boundary engineering) 
  {\em are restricted to the training data, and do {\bf not} affect the test  data.}
  \item There are many ways we could define ``actionable'' warnings. We use the definition that is consistent with past work~\cite{yang2021learning, yang2021understanding, wang2018there, kang2022detecting}. However, it can be argued that ``actionable'' refers to someone taking action (as demonstrated by an open issue being closed in a future commit, for example). Our definition then is a subset of the actionable ones according to this definition (since something could be actionable and some programmers choose to take no action). Another way to say that is that, by our definition, something is {\em definitely} actionable in the sense that in the historical record there is a clear record that some action was taken on it.
Perhaps it might be more precise to call these warnings ``the one that programmers choose to actually notice''-- which is what we take to be the essence of the prior work in this area~\cite{yang2021learning, yang2021understanding, wang2018there, kang2022detecting}.
\ei

 


\section{Background}
\label{sec:background}

\subsection{Static Code Analysis}
Automatic static analysis (SA) tools, such as Findbugs
are tools for   detecting bugs in source code,
without having to execute that code.
As they can find real bugs at low cost~\cite{thung2012extent,habib2018many}, they have been adopted in open source projects and in industry~\cite{ayewah2010google,sadowski2018lessons,beller2016analyzing,zampetti2017open,panichella2015would,vassallo2020developers}. For example, \citet{zheng2006value} discuss how a majority of the static analysis warnings were generated from a few patterns which lead to security vulnerabilities. Their study was based on defect data from over 3 million lines of C/C++ code.

However, as they do not guarantee that all warnings are real bugs, 
these tools produce false alarms. 
The large number of false alarms produced is a barrier to adoption~\cite{johnson2013don,ChristakisB16,tiganov2022designing,nachtigall2022large}; it is easy to imagine how developers will be frustrated by using tools that require them to inspect numerous false alarms before finding a real bug.
While this problem was raised by Johnson et al.~\cite{johnson2013don} nearly ten years ago,
a recent paper assessing the usability of static analysis tools~\cite{nachtigall2022large} confirmed that the challenge of too many false alarms remains an open problem.
While false alarms include spurious warnings caused by the over-approximation of possible program behaviors during program analysis, 
false alarms also refer to warnings that developers do not act on.
For example, developers may not think that the warning represents a bug (e.g. due to ``style'' warnings that developers perceive to be of little benefit) or may not wish to modify obsolete code.

The problem of addressing false alarms from static analysis tools has been widely studied~\cite{kharkar2022learning,muske2020techniques,kallingal2021validating,kim2022learning,kharkar2022learning,nam2019bug,croft2021empirical}.
There have been many recent attempts to address the problem. 
Some researchers have proposed new SA tools that use  more sophisticated, but costly, static analysis techniques (e.g. Infer~\cite{calcagno2015moving}, NullAway~\cite{banerjee2019nullaway}).
Despite their improvements, these tools still produce many false alarms~\cite{tomassi2021real}.
Another approach~\cite{nam2019bug} involves the   refinement of bug detection rules to avoid false positives, but requires manual analysis and design of the rules. 
Other attempts to prune false alarms include the use of test case generation to validate the presence of a bug  at the source code location indicated by the warning~\cite{kallingal2021validating}.
As generating test cases is expensive, these techniques may face issues when scaling up to larger projects, limiting their practicality.

\subsection{Early Results:  Wang et al., 2018}

By framing the problem as a binary classification problem, 
  machine learning techniques can identify actionable warnings 
  (allowing us to prune false alarms)~\cite{hanam2014finding,heckman2008establishing,liang2010automatic,ruthruff2008predicting,wang2018there,yang2021learning,yang2021understanding}.
These techniques use features extracted from code analysis and metrics computed over the code and warning's history in the project.

\begin{figure}[t]
  
  \includegraphics[width=\columnwidth]{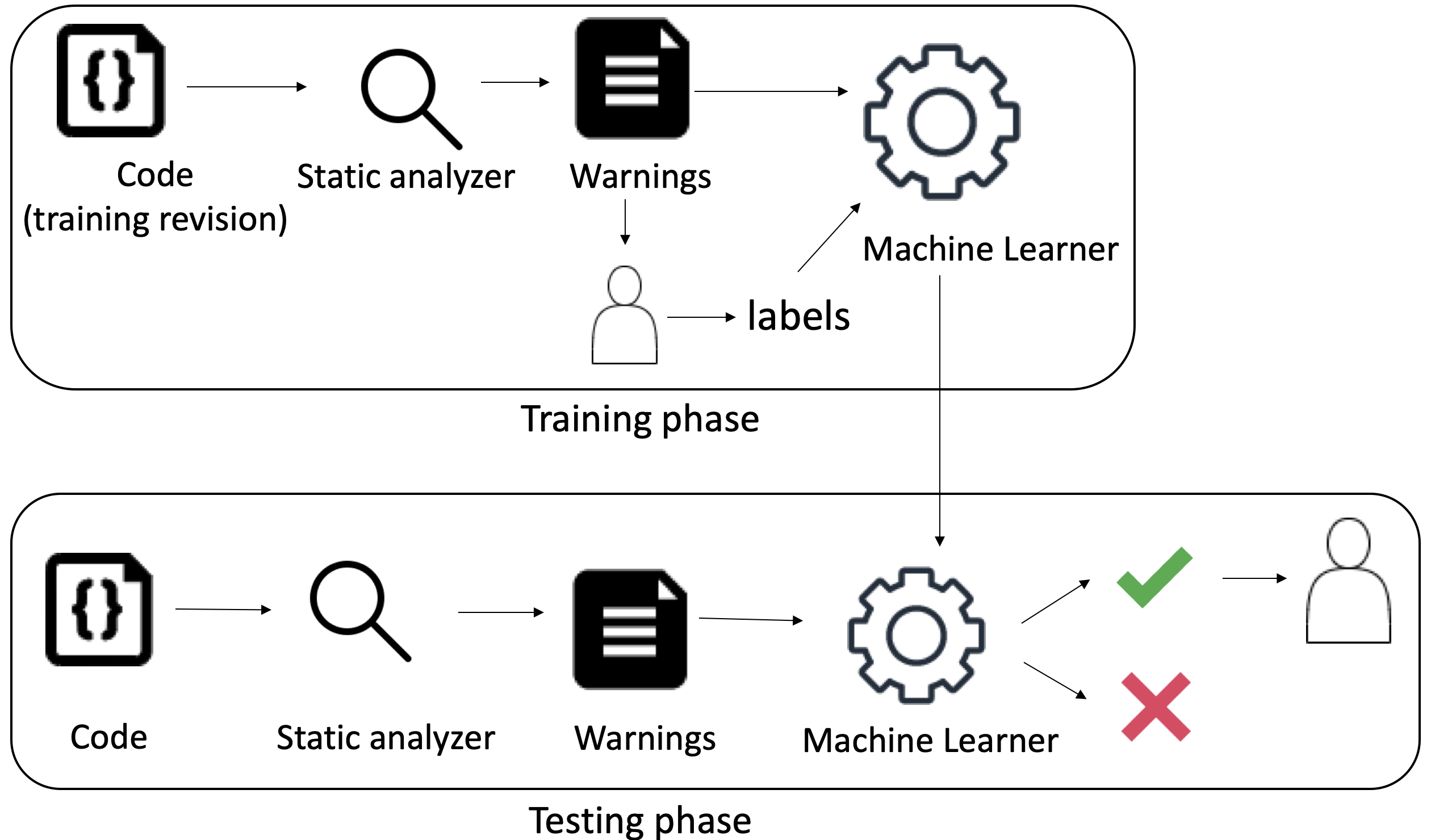}
  \caption{To detect actionable warnings, a   learner is trained   on warnings from a training revision. Each warning is annotated with a label. When deployed on the latest revision, only warnings classified as actionable warnings by the machine learner are presented to the developers. }
      \label{fig:workflow}
  \centering
  \end{figure}

Figure \ref{fig:workflow} illustrates this process. 
A static analyzer is ran on a training revision and the warnings produced are labelled. 
When applied to the latest revision, only warnings classified as actionable warnings by the machine learner are presented to the developers.

To assess proposed machine learners, datasets of warnings produced by Findbugs have been created.
As the ground-truth label of each warning is not known, a heuristic was applied to infer them.
This heuristic compares the warnings reported at a particular revision of the project against a revision set in the future.
If a warning is no longer present, but the file is still present, then the heuristic determines that the warning was fixed. 
As such, the warning is actionable.
Otherwise, if the warning is still present, then the warning is a false alarm.

 Wang et al.~\cite{wang2018there} ran a systematic literature review to collect  and analyze 100+ features proposed in the literature, categorizing them into 8 categories.
To remove ineffective features, they performed a greedy backward selection algorithm.
From the features, they identified a set of features that     offered effective performance.

\subsection{Further Result: Yang et al., 2021}

Yang et al.~\cite{yang2021learning} further analyzed the   features using the data collected by Wang et al.~\cite{wang2018there}. 
They found that all machine learning techniques were effective and performed similarly to one another. 
Their analysis revealed that the intrinsic dimensionality of the problem was low; 
the features used in the experiments were more verbose than the actual attributes required for classifying actionable warnings.
This motivates the use of simpler machine learners over more complex learners.
From their analysis, SVMs were recommended for use in this problem, as they were both effective and can be trained at a low cost.
In contrast, deep learners were effective but more costly to train.

For each project in their experiments, 
one revision (training revision) was selected for extracting warnings for training the learner, 
and another revision (testing revision) set chronologically in the future of the training revision is selected for extracting warnings for evaluating the learner. 
This simulates a realistic usage scenario of the tool, where the learner is trained using past data before developers apply it to  another revision of the source code.

\subsection{Issues in Prior Results: Kang et al., 2022}
Subsequently, Kang et al.~\cite{kang2022detecting} replicated  the Yang et al.~\cite{yang2021learning} study
to find subtle methodological issues in the     Wang et al. data~\cite{wang2018there} which   led to overoptimistic results.

Firstly, Kang et al. found data leakage where the information regarding the warning in the future, used to determine the ground-truth labels,  leaked into several features.
Five features (warning context in method, file, for warning type, defect likelihood, discretization of defect likelihood) measure the ratio of actionable warnings within a subset of warnings (e.g. warnings in a method, file, of a warning type). 
To determine if a warning is actionable, the ground-truth label was used to compute these features, leading to data leakage.
Kang et al.   reimplemented the features such that they are computed using only historical information, without reference to the ground truth determined from the future state of the projects.
As only the features were reimplemented, the total number of training and testing instances remained unchanged.

Secondly, they found many warnings appearing in both the training and testing dataset. 
As some warnings remain in the project at the time of both the training and testing dataset, the model has access to the ground-truth label for the warning at training time.
Kang et al. addressed this issue by removing warnings that were already present during the training revision from the testing dataset, ensuring that the learner does not see the same warning in both datasets.
After removing these warnings, the number of warnings in the testing revision decreased from
15,695 to 2,615.

\begin{table}[t]
    \centering
    \caption{Evaluation metrics based on TP (true positives); TN (true negatives);
    TP (true positives) and FP (false positives)}
    \label{tab:metrics}
      \rowcolors{2}{white}{gray!15}
    \begin{tabular}{rp{5cm}}
        \toprule
        \textbf{Evaluation Metric} & \textbf{Description}  \\
        \midrule
        Precision & $\frac{\text { TP } }{\text {TP}+\text {FP}}$ \\ 
    
        AUC  & area under the receiver operating
characteristics curve (the true positive
rate against the false positive rate) \\ 
   
        False alarm rate & $\frac{\text { FP } }{\text {FP}+\text {TN}}$\\
      
        Recall &  $\frac{\text { TP } }{\text {TP}+\text {FN}}$ \\
        \bottomrule
    \end{tabular}
\end{table}

\begin{table}[!t]
  \centering
  \caption{The predictors reported by Kang et al. did not perform well on the repaired data. In this table, {\em lower} false alarms are better while {\em higher} precisions, AUC, and recall are {\em better}. }
  \label{tab:initial_svm}
  \rowcolors{2}{white}{gray!15}
  \begin{tabular}{rrrrr}
 
 \toprule
                 
 \textbf{Dataset}           &  \textbf{Precision} & \textbf{AUC} & \textbf{False alarm rate} & \textbf{Recall}          \\
  \midrule 
  cassandra & 0.67 & 0.33 & 0.25 & 0.67 \\
   commons & 0.67 & 0.52 & 0.57 & 0.62 \\
   
  lucene-solr & 0.56 & 0.70  & 0.36 &  0.71\\
  
  maven & 0.52 & 0.41 & 0.19 & 0.32 \\
  jmeter & 0.50 & 0.36 & 0.14 & 0.17 \\
  
  tomcat & 0.52 & 0.41 & 0.19 & 0.32 \\
  
  derby & 0.20 & 0.64 & 0.12 & 0.08 \\
 
  ant & 0.00 & 0.00 & 0.00 & 0.00 \\
  \bottomrule
  \end{tabular}
  
\end{table}

Next, Kang et al. analyzed the warning oracle, based on the heuristic comparing warnings at one revision to another revision in the future, used to automatically produce labels for the warnings in the dataset. 
After manual labelling of the actionable warnings, Kang et al. found that only  47\% of warnings automatically labelled actionable were considered by the human annotators to be actionable.
This indicates that the heuristic employed as the warning oracle is not sufficiently reliable for automatically labelling the dataset. 

\begin{figure}[!b]
 \noindent {\footnotesize \begin{tabular}{c@{}c@{}c}
\includegraphics[height=.8in]{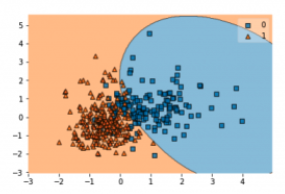}&
\includegraphics[height=.8in]{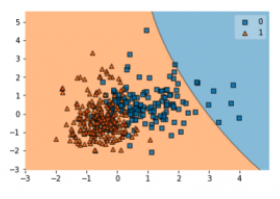}&
\includegraphics[height=.8in]{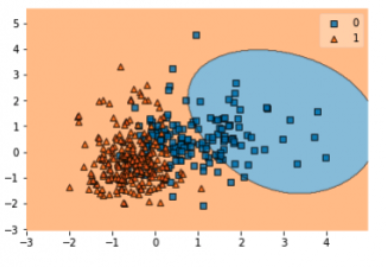}\\
$C=0.1$&$C=0.1$&$C=0.02$\\ 
$\gamma=0.1$&$\gamma=0.08$&$\gamma=0.1$\\
$\mathit{acc}=90\%$& $\mathit{acc}=64\%$&  $\mathit{acc}=81\%$ 
\end{tabular}}

\caption{The $C$  parameter of a radial basis function alters
the shape of the hyperspace boundary.  {\em Acc} is accuracy which is the ratio of true positives plus true negatives divided by a SVM making predictions across that boundary. Example from~\cite{kumar20}.}\label{cg}
\end{figure}

Kang et al. manually labelled 1,357 warnings. After filtering out duplicates and  uncertain labels, a dataset of 768 warnings remained.
On this dataset, Kang et al. again applied  off-the-shelf SVM models, assessing them with the evaluation metrics listed in Table \ref{tab:metrics}.

For their reasoning, 
Kang et al. used  the learners recommended by prior work;
i.e. radial bias  SVMs.
The results of the SVM are shown in Table \ref{tab:initial_svm}.
Those results are hardly impressive:
\bi
\item Median precisions barely more than 50\%;
\item Very low median AUCs of 41\%;
\item Extremely low median recalls of 32\%.
\ei
That is to say, while Kang et al. were certainly correct
in their criticisms of the data used in prior work, based on their paper,
it is still an open issue about how to   generate good predictors for static code false alarms.



\section{Rethinking the Problem}\label{problem}

This section suggests that detecting actionable static code warnings is a ``bumpy'' problem
(defined below) and that such problems can not be understood by learners that use simplistic
boundaries between classes.  

The core task of any classification problem is the creation of a hyperspace
boundary that let us isolate what is most desired or most interesting. Different learners build their boundaries in different ways:
\bi
\item Simple decision tree learners can only build straight-line
boundaries. 
\item Neural networks can produce very complex and convoluted boundaries.
\item And
internal to   
Kang et al.'s support vector machine was a ``radial basis function''
that allowed those algorithms to build circular hyperspace
boundaries.  
\ei 
Boundaries can be changed by adjusting
the
parameters that control the learner. For example, in Kang et al.'s radial basis functions,
the $C$ regularization parameter is used to set the tolerance of the model to (some) classifications. By adjusting $C$, an analyst can change
the generalization error; i.e. the error when the model is applied to as-yet-unseen test data.

Figure~\ref{cg} shows how changes to $C$ can alter the decision
boundary between some red examples and blue examples. Note that each
setting to $C$ changes the accuracy of the predictor; i.e.
for good predictions, it is important to fit the shape of the decision boundary to the shape of the data. 

(Technical aside: while this example was based on SVM technology, the same line of argument applies to  any other classifier; i.e. changing
the control parameters of the learner also changes the hyperspace boundary 
found by that learner and, hence, the predictive prowess of that learner.)

We have tried applying    hyperparameter optimization to   $C$ in a failed attempt to improve that performance (see the C1 results of Table \ref{tab:results}). 
From that failed experiment, we conclude that   however  $C$ works
for radial bias functions, they do not work well enough to fix
the unimpressive predictive performances -- see Table~\ref{tab:initial_svm}.

\begin{figure}[!t]
   \begin{center}\includegraphics[width=.3\textwidth]{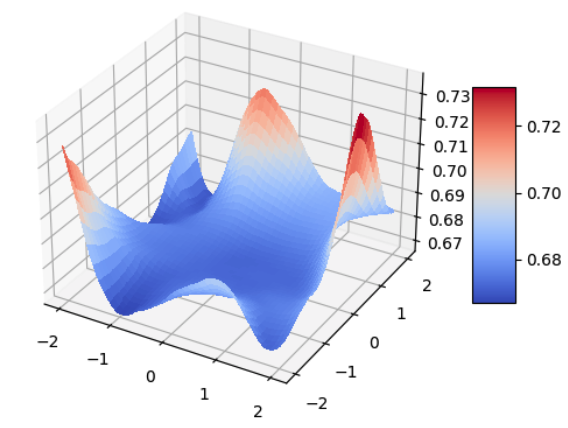} \end{center}
 
    \caption{Error landscape in the TOMCAT data    before    applying the methods of this paper. In the plot, the larger the vertical axes, the greater the loss value.
    Later in this paper,
    we will show this plot again, after it has been smoothed via the   methods of  \S\ref{rx}
    (see Figure~\ref{fig:gain} and Table~\ref{tab:fuzzy1}).}
    \label{fig:loss}
\vspace{-10pt}
\end{figure}

Why do radial bias SVMs fail in this domain?  Our  conjecture is that the hyperspace boundary dividing the static code examples (into false positives and others)
is so ``bumpy''\footnote{``Bumpy'' data 
  contain complexities such
as many local minima, saddle points, very flat regions,
and/or widely varying curvatures. For example, see  Figure \ref{fig:loss}.}
that the kinds of shape changes seen
in   Figure~\ref{cg} can never adequately model those  examples.

To test that conjecture, we first    checked for ``bumpiness" using a technique
from \citet{li2018visualizing}. That technique
visualizes the ``error landspace'' 
(i.e. how fast small changes in the independent variables altered the error
estimation).
For our TOMCAT data, Li et al.'s methods resulted in Figure \ref{fig:loss}.
There, we see a ``bumpy'' landscape
with several multiple local minima. 

Having confirmed that our data is ``bumpy'', our second step was to look  for ways to reduce that bumpiness. Initially, we attempted to use
  neural nets since that kind of learner is meant
to be able to handle complex hyperspace boundaries~\cite{WittenFH11}.
As discussed in \S\ref{sec:results}, that attempt failed even after trying
several different architectures such 
as
feedforward networks, CNN,  and CodeBERT~\cite{rumelhart1986learning,habib2018many,vaswani2017attention}
(with and without tuning learner control parameters).

Since standard neural net technology failed, we tried several   manipulation techniques for the training process, described in the next section.  

\section{  Treatments}\label{rx}

 This section discusses a framework that holds
 operators for   treating the data in order to adjust  the decision boundary (in different ways for different parts of the data). For the purposes of illustration and experimentation,
 we offer operational examples for each part of the framework:
 \bi
 \item  SMOTE for instance engineering;
 \item  SMOOTH for label engineering;
 \item   GHOST for boundary engineering;
 \item   DODGE for parameter engineering.
 \ei
 Before presenting those parts we note here that the framework is more than just those four treatments. 
 As SE research matures, we foresee that our framework will become a workbench   within which researchers replace some/all of these treatments with more advanced options.
 
 That said, we have some evidence that  SMOTE, SMOOTH, GHOST, DODGE are useful:
 \bi
  \item
The ablation study of \S\ref{rigg}  shows that removing any one of these treatments leads to worse performance.
 \item
All these treatments are  very fast: sub-linear time for SMOTE and SMOOTH, linear time for GHOST, and DODGE is known to be orders of magnitude faster than other hyperparameter optimizers~\cite{agrawal2019dodge}.
 \ei

\subsection{Instance Engineering (via SMOTEing)}
To remove the ``bumpiness'' in data like Figure~\ref{fig:loss}, we need to 
pull and push the decision boundaries between different classes into a smoother shape.
But also, unlike simplistic $C$ tuning available in radial  SVMs, we want that process to perform differently
in different parts of the data.

One way to adjust the decision boundary in different parts of the data is 
to add (or delete)  artificial examples around each example $X$. This builds a little ``hill'' (or valley) in the local region.
As a result, in that local region,
 it becomes more (or less)
certain that all predictions which reach the same conclusion as $X$. In effect, adding/deleting
examples pushes the decision boundary away (or, in the case of deletions, pulls it closer).
SMOTE \cite{chawla2002smote} is one instance engineering technique that:
\bi
\item Finds  five nearest neighbors
to   $X$ with the same label;
\item  Selects one at random;
\item Creates a new example $R$, with the same label as $X$ at some random point
between $X$ and $R$.
\ei

\subsection{Label Engineering (via SMOOTHing)}

SMOTE has seen much success in recent SE papers as a way to improve predication efficacy~\cite{agrawal2018better}. 
But this technique makes a {\em linearity} assumption that all the data around $X$ is correctly labelled
(in our case, as examples of actionable or unactionable static code warnings).
This may not be true.
 \citet{cordeiro2020survey} and recent SE researchers~\cite{frugal, debtfree, 9064604, jitterbug} note that  noisy labels can occur when human annotators are present~\cite{mcnicol2005primer} or those humans  have divergent opinions about the labels~\cite{barkan2021reduce, ma2019blind}. Although our labels were re-checked by the authors of \citet{kang2022detecting}, our ablation study (below) reports that it is best to apply some
 mitigation method for 
 poorly labelled examples. For example, in this work we applied the following SMOOTHing   operator where data is assigned labels using multiple near neighbors. This has the effect of removing outliers in the data. 
\BLUE \respto{2a3.1} At least within the domain of static analysis warning data, SMOOTHing, by reducing noise in the training labels, SMOOTH compensates for misclassification accuracy. \BLACK
 
 Our SMOOTH operator works as follows:
 \bi
 \item
 Given $n$ training samples (and therefore, $n$ labels), we keep $\sqrt{n}$ at random and discard the rest.
 \item
 Next, we use  a KD-tree to recursively sub-divide the remaining data into leaf clusters of   $\sqrt[4]{n}$ nearest neighbors. Within each leaf, all examples are assigned a label that is the  mode of the labels in that leaf.
 \ei
 One interesting and beneficial side-effect of SMOOTHing is that we   make conclusions on our test data using just 10\% of the training data. 
  By reducing  the labelling required to make conclusions, SMOOTHing
 offers a way to help future studies avoid the problems reported by Kang et al.~\cite{kang2022detecting}:
 \bi
 \item   One of the major finding of the Kang et al. study was that earlier work~\cite{yang2021learning}  had mislabelled much of its data. From that study, we assert that it is important for analysts to spend more time checking their labels.
 We note that there are many other ways to reduce the labels required for supervised learning. 
 \item SMOOTHing reduces the effort required for that checking
 process (by a factor of ten).
 \ei
 As an aside, we note that SMOOTHing
 belongs to a class of algorithms
 called {\em semi-supervised learning}~\cite{frugal, debtfree}  that try to   make conclusions using as few labels as possible.
The literature on semi-supervised learning is voluminous \cite{berthelot2019mixmatch, fairssl, kingma2014semi,  zhai2019s4l, zhu2005semi} and so, in the theory, there could be many other better ways to perform label engineering.
This would be a productive area for future research.    But for now, the ablation
  study (reported below) shows that   SMOOTHing is useful
  (since removing it degrades predictive performance).

\subsection{Boundary Engineering (via GHOSTing)}

As defined above, instance and label engineering do not reflect on the quality of data in the local region.

To counter that, this study employs a boundary method called ``GHOSTing'', recently developed and applied to software defect prediction  by    \citet{yedida2021value}.
 Boundary engineering is different to label and instance engineering since it adjusts the frequency of different classes
 in the local region (while the above typically end up repeating the same label for a particular locality). Hence, in that region,
 it changes the decision boundary.

GHOSTing addresses class imbalance issues in the data.   When an example with one label
is surrounded by too many examples of another label, then the signal associated with
example can be drowned out by its neighbors
To fix this,
for a two-class dataset $D$ with class $c_0$ being the minority, GHOSTing oversamples the class by adding concentric boxes of points around each minority sample. The number of concentric boxes is directly related to the class imbalance: higher the imbalance, more the number of boxes. Specifically, if $n$ is the fraction of samples in the minority class, then $\lfloor \log_2 (1/n) \rfloor$ boxes are added.
While the trivial effect of this is to oversample the class (indeed, as pointed out by \citet{yedida2021value}, this \textit{reverses} the class imbalance), we note that the algorithm effectively builds a wall of points around minority samples.
This pushes the decision boundary away from the training samples, which is preferred since a test sample that is close to a training sample has a lesser chance of   being
misclassified due to the decision boundary being in between
them.

Our pre-experimental intuition was that  boundary engineering would replace the need to use instance engineering. However, as shown by our ablation study, for recognizing actionable static code warnings, we needed both tools. On reflection, we realized   both may be  necessary since while (a)~boundary engineering can help make local adjustments to the decision boundary, it can (b)~only work in regions where samples \textit{exist}; instance engineering can help fill in gaps in sparser regions of the dataset.

\begin{table}[!b]
    \centering
    \caption{List of hyper-parameters tuned in our study.
    CodeBERT is not shown
  in that table since, as mentioned in the text, this analysis lacked
  the resources required to tune such a large model.}
    \label{tab:hyperparams}

    \begin{tabular}{llp{2.2cm}}
        \toprule
        \textbf{Learner} & \textbf{Hyper-parameter} & \textbf{Range}  \\
       
 \rowcolor{gray!15}         Feedforward network & \#layers & $[2, 6]$ \\
  \rowcolor{gray!15}        & \#units per layer & $[3, 20]$ \\
        
       Logistic regression & Penalty & $\{l_1, l_2\}$ \\
        & C & $\{0.1,1,10,100\}$ \\
        
  \rowcolor{gray!15}     Random forest & Criterion & $\{$ gini, entropy $\}$ \\
   \rowcolor{gray!15}       & n\_estimators & $[10, 100]$ \\
        
       Decision Tree & Criterion & $\{$ gini, entropy $\}$ \\
        & Splitter & $\{$ best, random $\}$ \\
         
  \rowcolor{gray!15}     SVM & C & $\{0.1,1,10,100\}$ \\
   \rowcolor{gray!15}       & Kernel & $\{$sigmoid, rbf, polynomial $\}$ \\
   
   CNN & \#convolutional blocks & [1, 4] \\
   & \#convolutional filters & \{4, 8, 16, 32, 64\} \\
   & Dropout probability & (0.05, 0.5) \\
   & Kernel size & \{16, 32, 64\} \\
        \bottomrule
    \end{tabular}
\end{table}

\subsection{Parameter Engineering (via DODGEing)}

We noted above that different learners
generate different hyperspace
boundaries (e.g. decision learners generate
straight-line borders while SVMs with radial bias functions generate circular borders). Further, once a learner
is selected, then as seen in Figure \ref{fig:loss}, 
it is possible to further adjust
a border by altering the   control parameters of
 that learner (e.g. see Figure~\ref{cg}).  We call this adjustment
 {\em parameter engineering}.
 
 Parameter engineering is like a scientist probing some phenomenon. After the data is divided into training and some
 separate test cases, parameter engineering algorithms conduct experiments on the training data looking for parameter settings
 that improve the performance of a model learned and assessed
 on the training data. Once some conclusions are
 reached about what parameters are best, then these are applied
 to the test data. Importantly, the parameter engineering  should only use the training    data for its investigations (since otherwise, that
 would be a threat to the external validity of the conclusions).

 Parameter engineering executes within the
 space of control parameters of selected learners. 
   These learners have the internal parameter space shown 
  in Table~\ref{tab:hyperparams}. 
We selected this range of learners using the following rationale:
 \bi
 \item
   In order to   compare our new results to   prior work by Kang et al. ~\cite{kang2022detecting}, we use the 
   Kang et al. {\em   SVMs}
  with the radial basis kernel and balanced class weights.
\begin{table}[!t]
    \centering
    \caption{Neural net architectures used
    in this study.}
    \label{tab:nnhere}
    
      \small
    \begin{tabular}{p{8.5cm}}
        \toprule
    \rowcolor{gray!15}
     {\em Feedforward networks}
     These are   artificial neural networks, comprising an acyclic graph of nodes that process input and produce an output. These dates back to the 1980s, and the parameters of these models are learned via backpropagation \cite{rumelhart1986learning}. 
       These networks have $\mathcal{O}(10^3)-\mathcal{O}(10^4)$ parameters. 
     For these networks, we used the    ReLU  (rectified linear activation)    function ($f(x) = \max (0, x)$).
     This is a  piecewise linear function that will output the input directly if it is positive, otherwise, it will output zero.  
        ~\\
        A {\em convolutional neural net} (CNN) is a structured neural net where the first several layers are sparsely connected in order to process information (usually visual).
        CNN is an example of an  
        {\em deep learner} and  are much larger than
        feedforward networks (these may   span $\mathcal{O}(10^5)-\mathcal{O}(10^7)$ parameters).
        Optimizing an CNN is a very complex task (so many parameters) so
        following   advice from the literature~\cite{ioffe2015batch,srivastava2014dropout}, we used the following architecture. Our CNNs had multiple ``convolutional blocks'' defined as follows:
\begin{enumerate}
    \item ReLU activation
    \item Conv (with ``same'' padding)
    \item Batch norm \cite{ioffe2015batch}
    \item Dropout \cite{srivastava2014dropout} 
\end{enumerate}

We note that this style of building convolutional networks, by building multiple ``convolutional blocks'' is very popular in the CNN literature \cite{krizhevsky2012imagenet,lecun1989backpropagation}. 
Our specific design of the convolutional blocks was based on a highly voted answer on Stack Overflow \footnote{\url{https://stackoverflow.com/a/40295999/2713263}}.

Note that with that architecture there is
still room to adjust the ordering of the blocks-- which is what we adjust when we tune our CNNs.\\   \rowcolor{gray!15} 
CodeBERT~\cite{feng2020codebert} is a {\em transformer-based}  model that been   pre-trained model using millions of examples
 from contemporary programming languages such as Python, Java, JavaScript, PHP, Ruby, and Go. 
Such transformer models
are those based on the ``self-attention'' mechanism proposed by \citet{vaswani2017attention}. CodeBERT
        is even large than CNN and can contain $\mathcal{O}(10^8)-\mathcal{O}(10^9)$ parameters.
One advantage of such large models is that can 
 learn intricacies that are missed by smaller models.   \\\bottomrule
    \end{tabular}
\end{table}    
 \item
   In order to   compare our   work to    Kang et al. ~\cite{yang2021learning}, we used 
  a range of  {\em traditional learners}
  (logistic regression, random forests, and single
 decision tree learners); 
 \item
Also, we explored the various  {\em    neural net algorithms}
shown in  Table~\ref{tab:nnhere}
since these algorithms have a reputation
 of being able to handle complex decision boundaries~\cite{WittenFH11}. 
 In this textbook on {\em Empirical
     Methods for AI}, Cohen~\cite{cohen1995empirical} advises that supposedly more complex solutions should be compared to  a range of alternatives, including very
     simple methods. Accordingly, for neural nets,
     we used (a)~feedforward networks from the 1980s;
     (b)~the CNN deep learner used in much of contemporary SE
     analytics; and (c)~the state-of-the-art   CodeBERT model.
 \ei     
There are many algorithms  available for automatically
tuning these learning control parameters.  
As recommended by a prior study \cite{agrawal2021simpler}, we use Agrawal et al.'s  DODGE algorithm~\cite{agrawal2019dodge}. 
DODGE is based on early work by Deb et al. in 2005 that proposed
a ``$\mathcal{E}$-domination rule''~\cite{Deb05}; i.e. 
\begin{quote}
{\em 
If one setting to an optimizer yield results within $\mathcal{E}$
or another, then declare the region $\pm\;\mathcal{E}$ as ``tabu''
and search elsewhere.}
\end{quote}
A surprising result from  Agrawal et al.'s research was that
$\mathcal{E}$ can be very large. 
Agrawal et al. noted that if learners were run 10 times, each time using 90\% of the training data (selected at random), then they often exhibited a  standard deviation of 0.05 (or more) in their performance scores.  Assuming that performance differences less than ${\pm}2\mu$, are
statistically insignificantly different, then Agrawal reasoned that   
$\mathcal{E}$ could be as large as $4*.05=0.2$.
This is an important point.
Suppose we are trying to optimize
for two goals (e.g. recall and false alarm). Since those
measures have
the range zero to one, then
$\mathcal{E}=0.2$ divides the output space of those two goals divides into just a $5{\times}5=25$ regions. Hence, in theory,  DODGE could find good
optimizations after just a few dozen random samples to the space of possible configurations.

When this theoretical prediction was checked
experimentally of SE data, Agrawal~\cite{agrawal2021simpler}  found that DODGE with 
$\mathcal{E}=0.2$ defeated traditional single-point cross-over genetic algorithms as well as state-of-the-art optimizers
(e.g. Bergstra and Bengio's HYPEROPT algorithm~\cite{Bergstra12}\footnote{At the time of this writing (April 2022), the paper proposing HYPEROPT has 7,557 citations in
Google Scholar.}).  Accordingly, this study used DODGE~\cite{agrawal2021simpler, dodge_comparison} for its parameter engineering.

Our pre-experimental intuition was that DODGEing would
be fast enough to tune even the largest neural net model. This turned out not to be the case. The resources required
to adjust the CodeBERT model are so large that, for this
study, we had to use the ``off-the-shelf'' CodeBERT.

\BLUE
In this study, whenever we say we perform hyper-parameter optimization (or equivalently in this paper, parameter engineering), we mean we run DODGE for 30 iterations to maximize the difference between recall and false alarm rate. This simultaneously aims to maximize recall while minimizing false alarm rate, prioritizing both goals equally.
\BLACK

\section{Experimental Methods}
\label{sec:methods}

 \subsection{Data}
This paper tested the efficacy of instance, label,
boundary and parameter engineering using the revised
and repaired data from 
  Kang et al. paper~\cite{kang2022detecting}.

Recall that
Kang et al. manually labelled warnings from the same projects studied by Yang et al.~\cite{yang2021learning} to assess the level of agreement between human annotators and the heuristic.
The manual labelling was performed by two human annotators. 
One annotator is an undergraduate student, while the other is a graduate student with two years of industrial experience.
When the annotators disagreed on the label of a warning, they discussed the disagreement to reach a consensus. 
While they achieved a high level of agreement, achieving a Cohen's Kappa of above 0.8, 
manual labelling is costly, requiring human analysis of both the source code and the commit history of the code.
That said, considering the subsequent evolution of the source code allows the annotators to analyze each warning with a greater amount of context.
These labels are essential
since it removed closed warnings which are not actionable (e.g., the warnings may have been removed for reasons unrelated to the Findbugs warning). 

Two other filters employed by Kang et al. where:
\bi
\item
Unconfirmed actionable warnings were removed;
\item
False alarms were randomly sampled to ensure a balance of labels (40\% of the data were actionable) consistent with the rest of the experiments.
\ei
One of the complaints of the Kang et al. paper~\cite{kang2022detecting}  against
earlier work~\cite{yang2021learning} was that, for data that comes with some time stamp, it is inappropriate to use future data to predict past labels. To avoid that problem,
in this study, we sorted the Kang et al. data by   time stamps,
then used 80\% of the past data to predict the remaining 20\% future labels.

The Kang et al. data comes from eight projects and we analyzed
each project's data separately. The 80:20 train:test splits
 resulted in the train:test sets shown in Table~\ref{tab:data}
 (exception: for MAVEN, we split 50:50, since there are only 4 samples in total). 

\begin{table*}[t]
\footnotesize
  \centering
  \caption{The dependent variables used in this study. These features were identified in prior work~\cite{wang2018there,yang2021learning,yang2021understanding}. The ``Golden Features'' are in bold.}
  \label{tab:golden_features}
  
    \begin{tabular}{l|l| l}
      \toprule
 \textbf{Feature type}     & \textbf{Feature} & \textbf{Description} \\
 \midrule
 & size context for warning type, method   & \\ 
 & size context in file, package  & \\ 
 
 & \textbf{warning context in method, file}, package  &  \\
 & \textbf{warning context for warning type}  & \\
 & fix, non-fix change removal rate. & \\
 Warning  & \textbf{defect likelihood for warning pattern} &  Features related to the warnings and other information  \\
 combination & variance of likelihood & (e.g. total number of warnings of each type, percentage  \\
 & defect likelihood for warning type & of actionable warnings)\\ 
 & \textbf{discretization of defect likelihood} & \\
 & \textbf{average lifetime for warning type}. & \\
 \midrule
 & method, file, package size. & \\
 & comment length   &\\
 & \textbf{comment-code ratio} & \\
 & \textbf{method depth}  &\\
  & \textbf{file depth} & \\
 Code  & method callers, callees &  Features related to the source file where the warning is  \\ 
 characteristics & \textbf{\# methods in file}, package & reported (e.g., the number of methods in the file) \\
 & \# classes in file & \\ 
 & \textbf{\# classes in package} & \\
 & indentation & \\ 
 & complexity &\\
 \midrule
 & \textbf{warning pattern} & \\
 & \textbf{warning type}  & \\
 Warning  & \textbf{warning priority}  & Features related to the warning (e.g., its priority) \\
 characteristics & warning rank, warnings in method, file &  \\
 & \textbf{package} & \\
 \hline
 & latest file, package modification & \\ 
 & file, package staleness &  \\
 File & \textbf{file age, creation} & Features related to the file where where the warning was reported  \\
 history & deletion revision & (e.g., creation date of the file) \\
 & \textbf{developers} & \\
 \hline
 & call name, class, \textbf{parameter signature} &  \\
 & \textbf{method visibility}  & \\
 & return type  & \\ 
 & new type, new concrete type & \\ 
 & operator & \\ 
 Code  & field access class, field  &  Features obtained through program analysis related to the source code \\ 
 analysis & catch & where the warning was reported  (e.g., if the method is public,  \\ 
 & field name, type, visibility, is static/final & protected, or private) \\ 
 & return type & \\ 
 & is static/ final/ abstract/ protected & \\ 
 & class visibility,is interface & \\
 \hline
 & added, changed, deleted, growth, total, &  \\
 & percentage of LOC in file (past 3 months)  & \\
  & LOC \textbf{added} , changed, deleted, growth&  \\
 Code  & total, percentage in file (last 25 revisions) &  features related to the revision history of the source code  \\
 history &  \textbf{added}, changed, deleted, growth, total, & where the warning was reported (e.g., number of \\ 
 & percentage of LOC in package  (past 3 months) & lines of code added in past 3 months) \\
 & added, changed, deleted, growth, total,  & \\
 & percentage of LOC in package (last 25 revisions) & \\
\hline
Warning  & \textbf{warning lifetime by revision}, by time &features related to the history of the warning in the project \\
history & warning modifications, open revision & \\
\midrule
File  & file type, name &  features related to the metadata of the file \\
Characteristics & package name & \\
\bottomrule 
\end{tabular}
\end{table*}

In this data, the dependent variables provide information about the warning, file, and source code that the warning is reported on. 
They are further categorized based on how they are obtained, e.g. through the history of the code.
Finally, the variables fall into eight broad categories.
They are summarized in Table \ref{tab:golden_features}.
Prior studies on static analysis warnings have worked with datasets with a wide range of actionable warnings. For example, Heckman and Williams~\cite{heckman2008establishing} experimented on 2 datasets, one of which had a percentage of actionable warnings of 89\%. 
Imtiaz et al.~\cite{imtiaz2019developers} experimented on several datasets, one of which had a percentage of 49.5\% of actionable warnings. 
Therefore, the percentage of actionable warnings in our experiments is consistent with some prior studies. 
 
Pre-experimentally, we were concerned that learning from the smaller data sets of Table~\ref{tab:data} would complicate
our ability to make any conclusions from this data.
That is, we needed to know:

\begin{formal}\noindent
\rqn{5} \textit{Are larger training sets necessary (for the task of recognizing actionable static code warnings)?}
\end{formal}

This turned out not to be a critical issue.
As shown below, the performance patterns   
in our experiments were   stable across all the six smaller data sets used in this study.



Technical aside: In other papers, we have run repeated trials with multiple 80:20
splits for training:test data. 
This was not here since some of our data sets are too small (see the first few rows of  Table~\ref{tab:data})
that any reduction in the training set size might disadvantage the learning process. 
Hence, the external validity claims of this paper come from patterns seen in eight different software projects.

\subsection{Models} 
\label{sec:models}

In this section, we discuss the models used by our approach. Briefly, we use feedforward networks, which are neural networks where each layer is fully connected. We do not use more modern approaches such as batch normalization \cite{ioffe2015batch} or dropout \cite{srivastava2014dropout}; as pointed out in the original GHOST paper \cite{yedida2021value}, the work of \citet{montufar2014number} shows that by setting the number of hidden layers to at least the number of inputs, the lower bound of the number of piecewise linear regions of the decision boundary is non-zero, and therefore (more likely to be) non-trivial. 

Although there are practical challenges when optimizing such basic models that are overcome by recent advances \cite{santurkar2018does, li2018visualizing}, we use the approach of GHOST, instead relying on hyper-parameter optimization. The defect prediction study of the original paper showed that hyper-parameter optimization along with weighted loss functions and ``fuzzy sampling'' (in this paper, we refer to the combination of these techniques as ``GHOST'' for simplicity) suffices to achieve state-of-the-art results--this paper shows that result extends to the task of demarcating actionable static code warnings as well.

\BLUE 
\subsection{LIME}

\respto{1a1.1}
It is worth looking inside the models at this point instead of treating them as black boxes. In an effort to understand the most important variables, we used LIME \cite{ribeiro2016should}, a standard explanation algorithm. However, this pursuit was unsuccessful for several reasons:
\bi\item
Some of our data sets are very small and, for such small data sets, LIME reports that all features are unimportant.
\item
All our data sets have different attributes (e,g, who made a comment, what part of the code they commenting on, etc). So even though we could generate results with high performance values, we could not find common patterns across the different data sets. 
\ei
We conjecture that static code warning classification is a hard problem with a bumpy decision boundary that cannot be characterized by (e.g.) LIME's simple linear models. Rather, we  may need some hyper-dimensional inferred description, which is why we see low prediction scores in Table~\ref{tab:initial_svm}  and much higher scores when we apply neural technology. 

\BLACK

\begin{table}[!t]
\centering
    \caption{Summary of the data distribution}
    \label{tab:data}
    \rowcolors{2}{white}{gray!10}
\begin{tabular}{r|rrrr}
\toprule
Project & \# train & \# labels & imbalance\% & \# test \\
\midrule
 maven & 2 & 1 & 33 & 1 \\
 cassandra & 9 & 4 & 38 & 4 \\
 jmeter & 10 & 4 & 43 & 4 \\
 commons & 12 & 5 & 59 & 5 \\
 lucene-solr & 19 & 5 & 38 & 6 \\
 ant & 22 & 6 & 36 & 7 \\
 tomcat & 134 & 13 & 41 & 37 \\
 derby & 346 & 20 & 37 & 92 \\
 \midrule
 total & 554 & 58 & & 156 \\
 \bottomrule
\end{tabular}
\end{table}
\begin{table*} 
    \centering
    \caption{Design of our ablation study. 
    In the learner choice column, \textbf{F} = feedforward networks, \textbf{T} = traditional learners, \textbf{C} = CNN, \textbf{B} = CodeBERT. 
    }
    \label{tab:treatments}
    \footnotesize
    \rowcolors{2}{white}{gray!10}
    \adjustbox{max width=\linewidth}{
    \begin{tabularx}{\linewidth}{r|lllll|rL}
        \toprule
        & \multicolumn{5}{c|}{Engineering decisions}&\\\cline{2-6}
        Treatment & \begin{turn}{70} Boundary    \end{turn} & \begin{turn}{70} Label   \end{turn} & \begin{turn}{70} Learner  \end{turn} & \begin{turn}{70}  Parameter   \end{turn} & \begin{turn}{70} Instance    \end{turn} & \% Labels  & Description \\
        \midrule
        A1   & \tick & \tick & F & \tick & \tick & 10 & Our recommended method \\
        A2 & \tick & \tick & F & \tick & & 10 & A1 without instance engineering (no SMOTE) \\
        A3 & \tick & \tick & F & & \tick & 10 & A1 without hyper-parameter engineering (no DODGE) \\
        A4 & & \tick & F & \tick & \tick & 10 & A1 without boundary engineering (no GHOST) \\
        A5 & \tick & & F & \tick & \tick & 100 & A1 without label engineering (no SMOOTH). From TSE'21~\cite{yedida2021value} \\
        A6 & \tick & & T & \tick & \tick & 100 & A1 without label engineering, replacing feedforward with traditional learners \\
        A7 & \tick & \tick & T & \tick & \tick & 10 & A1 replacing feedforward with traditional  learners \\
        B1 & & \tick & T & \tick & \tick & 10 & A1 without boundary engineering, replacing feedforward with traditional learners \\
        B2 & & \tick & C & \tick & \tick & 10 & A1 without boundary engineering, replacing feedforward with CNN \\
        C1 & & & T & \tick & \tick & 100 & A1 without boundary engineering or label engineering, replacing feedforward with traditional learners \\
        C2 & & & C & \tick & \tick & 100 & A1 without boundary engineering or label engineering, replacing feedforward with CNN \\
        D1 & & & T & & \tick & 100 & Setup used by the \citet{yang2021learning} and \citet{kang2022detecting} studies. \\
        CodeBERT & & & B & & & 100 & CodeBERT without modifications \\\bottomrule
    \end{tabularx} }
\end{table*}
\subsection{Experimental Rig}\label{rigg}
 This study explores:
 \bi
 \item
   $N=4$ pre-processors 
 (boundary, label, parameter, instance)
 that could be mixed in $2^4=16$ ways.
 \item
 Six  traditional learners:  logistic regression, decision trees,
 random forests,
 SVMs (with 3  basis functions);
 \item 
 Three neural net architectures: CNN, CodeBERT, feedforward networks;
 \ei
  To clarify  the reporting of these $16\times(6+3) = 144$ treatments, we made the following decisions.
  Firstly, 
  when reporting the results of the traditional learner, just show the results of the one that beat the other traditional learners   (which, in our case, was typically random forest or logistic regression).
 
Secondly, we 
  do not apply pre-processing or parameter engineering
  on CodeBERT. This decision was required, for pragmatic reasons. Due to  the computational cost of training
  that model,  we could  only  run   off-the-shelf CodeBERT.
 
 Thirdly, 
  rather than explore all 16 combinations of use/avoid
  different pre-processing, we  ran the {\em ablation study}
  recommended in Cohen's
  {\em Empirical
     Methods for AI} textbook~\cite{cohen1995empirical}.
    Ablation studies let us explore some combination of
     $N$ parts can be assessed in time $O(N)$, not $O(2^N)$.
     Such   ablation studies work as follows:
     \bi
     \item Commit to a preferred approach, with $N$ parts;
     \item If removing any  part $n_i\in N$ degrades performance, then conclude that  all  $N$ parts are useful.
     \ei
 With these decisions, 
 instead of having to report on 144 treatments,
 we need only show the 13 treatments in the  ablation study of
Table~\ref{tab:treatments}.
In that table,
for treatments that use any of boundary  
or label or parameter or instance engineering,
we apply those treatments in the  order  recommended by  the original  GHOST paper~\cite{yedida2021value}. That paper
 found  that it could improve  recall by 30\% (or more) by multiple rounds of SMOTE + GHOST. As per that advice,
 A1  executes our pre-processors in the  order:
   \[ \mathit{smooth} \rightarrow \mathit{smote} \rightarrow \mathit{ghost} \rightarrow \mathit{ghost} \rightarrow \mathit{smote} \rightarrow  \mathit{dodge} \]
   
The rationale for this approach
 is as follows.
Learning can be divided into three stages: preprocessing, ``in-processing'', and post-processing.  This paper does not explore post-processing (and that might be a useful direction for future work).
The crux of this paper (and the original GHOST paper) is that we should focus more than we currently do on pre- and in-processing. As such, we start the pre-processing with SMOOTH, which has the effect of eliminating outliers in the data. We then balance classes using SMOTE. After that, we applied a recommendation from the GHOST paper, i.e. it is best to run GHOST twice on the data (hence our call to GHOST -$>$ GHOST). It turns out that we  should not pass those outputs directly to the   learner, since we want a robust decision boundary.
Hence, we use another recommendation from the  GHOST paper which is 
\[\mathit{ghost}  \rightarrow \mathit{ghost}  \rightarrow  \mathit{smote}\]

\BLUE
All that said, it is an open issue if {\em other} ordering might be {\em more} useful. In this paper, we mote that
our ablation study reports no obvious problem with this ordering. But that is not to say that other orderings
might improve our results even further. We leave this matter for future work.
\BLACK


  
All the treatments labelled ``A'' (A1,A2,A3,A4,A5)  in Table~\ref{tab:treatments},  use the order shown above, perhaps
(as part of the ablation study) skipping over one or more the steps. 
We acknowledge that there are many possible ways to order the applications of our treatments, which is a matter we will for future work. For the moment,the ordering shown above seems useful (evidence: see next section).


\newcommand{\best}{\cellcolor{lightgray}}
\newcommand{\bad}{\cellcolor{pink}}

\begin{table*}[t]
\renewcommand{\baselinestretch}{.5}
{ 
   
    \caption{Results (8 datasets, 4 metrics. Pink shows   performance
  worse than A1 (our recommend method).}
    \label{tab:results}
   \scriptsize
    \begin{center}
    \begin{tabular}{l|llllllll|l|c}
    
        \toprule   
        \textbf{Treatment} & \textbf{maven} & \textbf{cassandra} & \textbf{jmeter} & \textbf{commons} & \textbf{lucene-solr} & \textbf{ant} & \textbf{tomcat} & \textbf{derby} & \textbf{median} &
        \textbf{\#cells better than A1}\\
        \midrule
        \multicolumn{10}{c}{PRECISION  ({\em better} results are {\em larger})} \\
        \midrule
A1                      & 1   & 1    & 1    & 1    & 0.8  & 1    & 0.79 & 0.72 & 1       & -      \\
A2                      & \bad 0   & \bad 0.25 & 1    & \bad 0.67 & 1    & 1    & \bad 0.68 & 0.73 & \bad 0.71  & 2        \\
A3                      & \bad 0.5 & \bad 0.25 & \bad 0.33 & \bad 0.2  & \bad 0.25 & \bad 0.33 & \bad 0.33 & \bad 0.4  & \bad 0.33 & 0          \\
A4                      & 1   & \bad 0.75 & 1    & 1    & 0.75 & 1    & 1    & \bad 0.75 & 1   & 2          \\
A5                      & 1   & 1    & 1    & 1    & 0.8  & 1    & \bad 0.72 & 0.84 & 1     & 1        \\
A6                      & 1   & 1    & \bad 0.5  & \bad 0.33 & \bad 0.67 & 1    & 0.85 & 0.89 & 0.87 & 2 \\
A7                      & 1   & \bad 0.5  & \bad 0.5  & 1    & 1    & \bad 0    & \bad 0.55 & \bad 0.42 & \bad 0.53  & 1        \\
B1 (DODGE)              & 1   & 1    & \bad 0    & \bad 0.5  & \bad 0    & \bad 0    & \bad 0.47 & \bad 0.59 & \bad 0.49  & 0        \\
B2 (CNN)                & \bad 0.5 & \bad 0    & \bad 0    & \bad 0.6  & \bad 0    & \bad 0.29 & \bad 0.51 & \bad 0.61 & \bad 0.4      & 0     \\
C1 (DODGE)              & 1   & 1    & 1    & \bad 0.33 & \bad 0.67 & \bad 0.67 & \bad 0.67 & 0.81 & \bad 0.74   & 1       \\
C2 (CNN)                & \bad 0.5 & \bad 0.5  & \bad 0.5  & \bad 0.6  & 0.83 & \bad 0.43 & \bad 0.4  & 0.73 & \bad 0.5     & 2      \\
D1                      & \bad 0   & \bad 0.5  & \bad 0    & \bad 0.6  & \bad 0    & \bad 0    & \bad 0.39 & \bad 0.39 & \bad 0.2     & 0      \\
CodeBERT & \bad 0.5 & 1    & \bad 0.8    & \bad 0.63  & \bad 0.6  & \bad 0  & \bad 0.41 & \bad 0.25 & \bad 0.55 & 0 \\
    \midrule 
    \multicolumn{10}{c}{AUC: TP vs. TN ({\em better} results are {\em larger})}   \\
    \midrule
A1                   & 1   & 1    & 0.83 & 1    & 0.75 & 1    & 0.68 & 0.57 & 0.92   & -       \\
A2                   & \bad 0   & \bad 0.5  & 0.75 & 0.83 & 0.63 & 0.75 & 0.6  & 0.7  & \bad 0.67   & 1        \\
A3                   & \bad 0.5 & \bad 0.5  & \bad 0.67 & \bad 0.5  & \bad 0.38 & \bad 0.55 & \bad 0.51 & 0.51 & \bad 0.51   & 0        \\
A4                   & 1   & 0.5  & 1    & 0.75 & 0.63 & 0.8  & 0.54 & 0.59 & 0.69     & 2     \\
A5                   & 1   & 0.67 & 1    & 0.88 & 0.75 & 0.9  & 0.67 & 0.78 & 0.83    & 2      \\
A6                   & 1   & 1    & 0.83 & 0.75 & 0.88 & 1    & 0.85 & 0.76 & 0.87 & 3 \\
A7                   & 1   & 0.83 & 0.83 & 1    & 0.75 & 0.5  & 0.59 & 0.62 & 0.79   & 1       \\
B1 (DODGE)           & 1   & 1    & 0.5  & 0.88 & 0.5  & 0.5  & 0.58 & 0.62 & 0.6   & 1        \\
B2 (CNN)             & \bad 0.5 & \bad 0.5  & \bad 0.5  & \bad 0.5  & \bad 0.5  & \bad 0.5  & \bad 0.5  & 0.62 & \bad 0.5         & 1  \\
C1 (DODGE)           & 1   & 1    & 1    & 0.75 & 0.88 & 0.9  & 0.8  & 0.76 & 0.89  & 4        \\
C2 (CNN)             & \bad 0.5 & \bad 0.5  & \bad 0.17 & \bad 0.5  & \bad 0.5  & \bad 0.5  & 0.63 & 0.82 & \bad 0.5  & 1         \\
D1                   & \bad 0.5 & \bad 0.17 & \bad 0.5  & \bad 0.5  & \bad 0    & \bad 0.38 & \bad 0.48 & \bad 0.47 & \bad 0.48    & 0      \\
CodeBERT & \bad 0.5 & \bad 0.56 & \bad 0.68 & \bad 0.53 & \bad 0.63 & \bad 0.48 & \bad 0.44 & 0.63 & \bad 0.54 & 1 \\
    \midrule
    \multicolumn{10}{c}{FALSE ALARM RATE ({\em better} results are {\em smaller})} \\
    \midrule
A1                     & 0 & 0    & 0    & 0    & 0.5  & 0    & 0.29 & 0.79 & 0  & -            \\
A2                     & 0 & 1    & 0    & 0.33 & 0    & 0    & 0.4  & 0.38 & 0.17  & 2        \\
A3                     & \bad 1 & \bad 1    & \bad 0.67 & \bad 1    & \bad 0.75 & \bad 0.4  & \bad 0.71 & 0.05 & \bad 0.73    & 1      \\
A4                     & 0 & 1    & 0    & 0    & 0.5  & 0    & 0    & 0.48 & 0    & 2         \\
A5                     & 0 & 0    & 0    & 0    & 0.5  & 0    & 0.57 & 0.41 & 0    & 1         \\
A6                     & 0 & 0    & 0.33 & 0.5  & 0.25 & 0    & 0.09 & 0.03 & 0.06 & 3 \\
A7                     & 0 & 0.33 & 0.33 & 0    & 0    & 0    & 0.17 & 0.44 & 0.09   & 3       \\
B1 (DODGE)             & 0 & 0    & 0    & 0.25 & 0    & 0    & 0.35 & 0.11 & 0    & 2         \\
B2 (CNN)               & \bad 1 & 0    & 0    & \bad 1    & 0    & \bad 1    & \bad 1    & 0.25 & \bad 0.63    & 2      \\
C1 (DODGE)             & 0 & 0    & 0    & 0.5  & 0.25 & 0.2  & 0.26 & 0.06 & 0.13    & 3      \\
C2 (CNN)               & \bad 1 & \bad 1    & \bad 1    & \bad 1    & \bad 1    & \bad 1    & 0.46 & 0.17 & \bad 1   & 1          \\
D1                     & 0 & \bad 1    & 0    & \bad 1    & \bad 1    & \bad 0.25 & \bad 0.77 & \bad 0.67 & \bad 0.72  & 0        \\
CodeBERT & \bad 1 & 0 & \bad 0.2  & \bad 1 & 0.25 & \bad 0 &  0.28 &  0.17 & \bad 0.23 & 2  \\
    \midrule
    \multicolumn{10}{c}{RECALL  ({\em better} results are {\em larger})} \\
    \midrule
A1                     & 1   & 1    & 0.67 & 1    & 1    & 1   & 0.65 & 0.94 & 1 & - \\
A2                     & \bad 0.5 & 1    & \bad 0.5  & 1    & \bad 0.25 & \bad 0.5 & 0.59 & \bad 0.77 & \bad 0.54  & 0     \\
A3                     & 1   & 1    & 1    & 1    & 0.5  & 0.5 & 0.75 & 0.07 & 0.88    & 2   \\
A4                     & 1   & 1    & 1    & \bad 0.5  & \bad 0.75 & \bad 0.6 & \bad 0.09 & \bad 0.67 & \bad 0.71   & 1     \\
A5                     & 1   & \bad 0.33 & 1    & \bad 0.75 & 1    & \bad 0.8 & 0.91 & 0.97 & 0.94     & 3   \\
A6                     & 1   & 1    & 1    & 1    & 1    & 1   & 0.79 & 0.55 & 1  & 2 \\
A7                     & 1   & 1    & 1    & 1    & 0.5  & 0   & 0.36 & 0.69 & 0.85     & 1  \\
B1 (DODGE)             & 1   & 1    & \bad 0    & 1    & \bad 0    & \bad 0   & \bad 0.5  & \bad 0.34 & \bad 0.42  & 0     \\
B2 (CNN)               & 1   & 0    & 0    & 1    & 0    & 1   & 1    & 0.49 & 0.75  & 1     \\
C1 (DODGE)             & 1   & 1    & 1    & 1    & 1    & 1   & 0.86 & 0.59 & 1 & 2          \\
C2 (CNN)               & 1   & 1    & 0.33 & 1    & 1    & 1   & 0.73 & 0.82 & 1 & 1 \\
D1                     & \bad 0   & \bad 0.33 & \bad 0    & 1    & \bad 0    & \bad 0   & \bad 0.73 & \bad 0.61 & \bad 0.17    & 0   \\
CodeBERT & 1   & \bad 0.33 & 0.67 &  1 & \bad 0.5 & \bad 0   & \bad 0.26  & \bad 0.25 &  \bad 0.42  & 0\\
    \bottomrule
    \end{tabular}
    \end{center}
}
\end{table*}

\begin{table*}[t]
\renewcommand{\baselinestretch}{.5}
{ 
   
    \caption{Our results using A1, with varying train/test ratios.}
    \label{tab:datasize}
   \scriptsize

    \begin{center}
    \begin{tabular}{l|llllllll|l}
    
        \toprule   
        \textbf{\% train} & \textbf{maven} & \textbf{cassandra} & \textbf{jmeter} & \textbf{commons} & \textbf{lucene-solr} & \textbf{ant} & \textbf{tomcat} & \textbf{derby} & \textbf{median} \\
        \midrule
        \multicolumn{10}{c}{PRECISION  ({\em better} results are {\em larger})} \\
        \midrule
        80\% & 1 & 1 & 1 & 1 & 0.8 & 1 & 0.79 & 0.72 & 1 \\
        60\% & 1 & 1 & 1 & 0.8 & 0.8 & 1 & 0.65 & 0.67 & 0.9 \\
        40\% & - & 0 & 0.86 & 0.8 & 0.82 & 0.67 & 0.65 & 0.65 & 0.67 \\
        20\% & - & 0 & 0.73 & 0.57 & 0.69 & 0.67 & 0.76 & 0.65 & 0.67 \\
    \midrule 
    \multicolumn{10}{c}{AUC: TP vs. TN ({\em better} results are {\em larger})}   \\
    \midrule
        80\% & 1 & 1 & 0.83 & 1 & 0.75 & 1 & 0.68 & 0.57 & 0.92 \\
        60\% & 1 & 0.63 & 0.8 & 0.9 & 0.75 & 0.83 & 0.55 & 0.56 & 0.78 \\
        40\% & - & 0.5 & 0.8 & 0.95 & 0.77 & 0.54 & 0.56 & 0.53 & 0.56 \\
        20\% & - & 0.5 & 0.69 & 0.71 & 0.59 & 0.61 & 0.7 & 0.54 & 0.61 \\
    \midrule
    \multicolumn{10}{c}{FALSE ALARM RATE ({\em better} results are {\em smaller})} \\
    \midrule
        80\% & 0 & 0 & 0 & 0 & 0.5 & 0 & 0.29 & 0.79 & 0 \\
        60\% & 0 & 0 & 0 & 0.2 & 0.5 & 0 & 0.52 & 0.81 & 0.1 \\
        40\% & - & 0 & 0.25 & 0.1 & 0.29 & 0.38 & 0.57 & 0.8 & 0.29 \\
        20\% & - & 0 & 0.5 & 0.25 & 0.56 & 0.73 & 0.36 & 0.55 & 0.5 \\
    \midrule
    \multicolumn{10}{c}{RECALL  ({\em better} results are {\em larger})} \\
    \midrule
        80\% & 1 & 1 & 0.67 & 1 & 1 & 1 & 0.65 & 0.94 & 1 \\
        60\% & 1 & 0.25 & 0.6 & 1 & 1 & 0.67 & 0.62 & 0.92 & 0.8 \\
        40\% & - & 0 & 0.86 & 1 & 0.82 & 0.46 & 0.69 & 0.86 & 0.82 \\
        20\% & - & 0 & 0.89 & 0.67 & 0.73 & 0.94 & 0.75 & 0.63 & 0.73 \\
    \bottomrule
    \end{tabular}
    \end{center}
}
\end{table*}

As to the specifics of the other treatments:
\bi
\item
Treatment A5 is the treatments from the TSE'21 paper  that   proposed GHOSTing~\cite{yedida2021value}.
\item Treatment D1 contains the treatments applied in prior papers by   Yang et al.~\cite{yang2021learning}  and Kang et al. ~\cite{kang2022detecting}.
\item Anytime we applied {\em parameter engineering},
this meant that some automatic algorithm (DODGE) selected the control parameters for the learners (otherwise, we just used the default off-the-shelf settings). 
\item Anytime we apply {\em label engineering}, we are only
used 10\% of the labels in the training data.
\item The last line, showing CodeBERT, has no pre-processing or tuning. As said   above, CodeBERT is so complex that we must  run it ``off-the-shelf''.
\ei

\section{Results}
\label{sec:results}

\newcommand{\ans}[1]{\underline{\textbf{Answer~#1:}}}

  \begin{table} 
  \caption{Median performance improvements seen after applying all the treatments A1 (defined in \S\ref{rx}); i.e. all of
{\em instance}, {\em label}, {\em boundary}  and {\em parameter} engineering.}
 \begin{center}
{\scriptsize  \begin{tabular}{|c|c|c|c|c|}\cline{3-5} 
    \multicolumn{2}{c|}{} &       &  From right-& \\ 
    \multicolumn{2}{c|}{} & From     &  hand-side & \\  
     \multicolumn{2}{c|}{} &   Table~\ref{tab:initial_svm}  &  of Table~\ref{tab:results} &  Improvement \\\hline
                             & precision   & 50 & 100 & 50\\ 
{\em higher} is {\em better} & AUC         &  41& 90 & 59\\ 
                             & recall      &  19 & 100 & 89\\\hline
{\em lower} is {\em better}  & false alarm & 32 & 0 & 32\\\hline
\end{tabular}} 
\end{center}
\label{summary}
\end{table}



The results of the Table~\ref{tab:treatments} treatments are shown in   Table~\ref{tab:results}
(and 
another brief summary is offered in
Table~\ref{summary}).
These results are somewhat
extensive
  so, by way of an overview, we offer the following summary tool.
 The cells shown in  \colorbox{pink}{pink} are those that are   worse than the A1 results 
  (and A1 is our recommended GHOST2 method). 
 Looking over those pink cells
 we can see that across our data sets and across our different measures, our recommend method (A1) is rarely outperformed by anything else. 
 
 (Technical aside: looking at the pink cells, it could be said that   A5 comes close to A1, but A5 loses a little on recalls).
 Nevertheless, we have strong reasons for recommending A1 over A5 since, recalling Table~\ref{tab:treatments},
 A5 requires a labelling for 100\% of the data. On the other hand A1, that uses label engineering,
 achieves its results using 10\% of the labels. This is important since, as  said in our introduction, 
one way to address, in part, the methodological problems
raised by Kang et al. GHOST2 makes its conclusions
using a small percentage of the raw data (10\%). That is,
to address issues of corrupt data found by Kang et
al., we say ``use less data'' and, for the data that is used,
``reflect more on that data''.
 
These  answer our research questions as follows.

\subsection*{ \rqn{1} For detecting actionable static code warnings,
what data mining methods should we recommend?}

Regarding feedforward networks versus, say,  traditional learners (decision trees, random forests, logistic regression and SVMs), the traditional learners all performed worse than the 
feedforward networks used in treatment A1 (evidence: compare treatments
A1 with A7 which use feedforward or traditional learners, respectively; there are four perfect AUCs
for feedforward networks in A1, i.e AUC=100\%, but only two for the A7 results).

As to why 
the 1980s style
feedforward networks worked better than newer neural net technology,  we note that feedforward networks run so fast
than it is easier to extensively tune them. Perhaps
 (a)~faster learning plus (b)~more tuning might lead to better results
that then non-linear modeling of an off-the-shelf learner.
This could be an interesting avenue for future work.

As to the value of {\em boundary, label, instance}
and {\em parameter} engineering,   in the ablation study, removing any of these increased the number of times A1 had larger performance scores.
For example, 
with {\em boundary engineering},
   A1
(that uses boundary engineering) generates more perfect
scores (e.g. AUC=100\%) than A4 (that does not use it).
Also, for recall, A1 always had larger or same scores
than  n A4 in 6/8 data sets. Similarly, A4 always suffers from a drop in AUC score.

As for {\em label engineering}, from A1 to A5,
specializing our data to just 10\% of
the labels (in A1) yields nearly the same precisions
which using 100\% of the data (in A5) in nearly all the AUC results. Moreover, the AUC score for A1 is perfect in 4/8 cases, while for A5, it is rarely the case.

As to {\em instance engineering},
without it the  
precision can crash to zero
(compare A1 to A2, particularly  the smaller data sets) while often leading to
lower recalls. The smaller datasets also see a decrease in AUC for A2.

Measured in terms of false alarm, these results strongly recommend 
{\em parameter engineering}. Without parameter engineering, some of those treatments could find too many static code warnings and hence suffer from
excessive false alarms (evidence: see the A3 false alarm results in nearly every data set). A1 (which used all the   treatments of \S\ref{rx}) had lower false alarm rates than anything else (evidence: we rarely see the dark blue A1 spike in the false alarm results). The  only exception to the observation that ``parameter engineering leads to lower false alarm results''
are seen in the   DERBY
data set. That data set turns out to be particularly tricky in that, nearly always, modeling methods  that achieved low false alarm rates on that data set
also had to be satisfied with much lower recalls.

One final point is that these results do not  recommend 
the use of certain widely used neural network
technologies such as CNN or CodeBERT for finding actionable
static code warnings.
CNN-based treatments (B2 and C2) suffer from low precision and AUC scores (see Table \ref{tab:results}).
Similarly, as shown Table~\ref{tab:results}, CodeBERT
often suffers from   low precision and poor false alarms
and (in the case of CodeBERT) some very low recalls indeed.

\BLUE \respto{2a4.1}
Finally, we also test how few training samples we can get away with before significant drops in performance. For this, we test our experimental setup by changing the train/test ratio:
\bi
\item Our original setup used 80\% for the training set and 20\% for the test set; we test 60/40, 40/60, and 20/80 splits. These results are shown in Table \ref{tab:datasize}.
\item
While there are within-dataset variations, the medians show the expected result: as the training size reduces, so does performance.
\ei
That said,  the drop in performance when using  60\% of samples instead of 80\% is rather small; the major drop in performance comes when we drop down to 40\%.  At 40\% and lower, there are too few samples to train on maven (which has 4 samples in total), so the learners throw exceptions--this is why we cannot compute metrics. 

We note that even at 60\%, the median number of labels we use is $\sqrt{0.6n}$ where $n$ is the median number of samples; this computes to 4 samples. Across all datasets, the minimum value of this is 2, and the maximum value is 17. However, this is not to say this is true for all datasets--we note that as shown in Figure \ref{fig:gain}, the loss landscape after applying our methods was very flat, making optimization easier. Our expectation is that for other datasets, if GHOST and DODGE together flatten the loss landscape to this extent, then learning from sub-10 samples should be possible.
\BLACK

\noindent  In summary:
\begin{formal}
\ans{1} To recognize actionable static code warnings, apply all the treatments of \S\ref{rx}. Also, 
spend most tuning faster feedforward neural nets  rather than trusting (a)~traditional learners or (b)~more recent
``bleeding edge'' neural net methods.
\end{formal}

\subsection*{\rqn{2} Does GHOST2's combination of instance, label, boundary
and parameter engineering, reduce the complexity of the
decision boundary?}

Previously, this paper argued
that reason for the poor performance
seen in prior was due to the complexity
of the data (specifically, the bumpy shape seen in  Figure~\ref{fig:loss}).
Our treatments of \S\ref{rx} were  designed to simplify that landscape. Did we succeed?

 \begin{table}
     
    \caption{Percent changes in \citet{li2018visualizing}'s
    smoothness metric, seen after applying the methods of this paper.}
    \label{tab:fuzzy1}\footnotesize
    \begin{center}
    \begin{tabular}{lr}
        \toprule
        \textbf{Dataset} & \textbf{\% change} \\
        \midrule
        
   \rowcolor{gray!15}       maven & 158.87 \\
        
        cassandra & 73.09 \\
        
        \rowcolor{gray!15}       jmeter & 55.53 \\
        tomcat & 36.34 \\
        \rowcolor{gray!15}       derby & 31.35 \\
        commons & 29.61 \\
         \rowcolor{gray!15}      ant & 24.78 \\
        lucene-solr & 16.46 \\
        \midrule
        \textbf{median} & \textbf{33.85} \\
        \bottomrule
    \end{tabular}
    \end{center}
\end{table}
Figure~\ref{fig:gain}
 shows the landscape in  TOMCAT after
 the   treatments of \S\ref{rx}
 were applied.
 By comparing this figure with
      Figure~\ref{fig:loss},
      we can see that our treatments
      achieved the desired goal
      of removing the ``bumps''.
      
 As to the other data
      sets,
      \citet{li2018visualizing}
      propose a ``smoothness'' equation
      to measure a data set's
      ``bumps''.
      Table~\ref{tab:fuzzy1}
      shows the percentage 
      change in that smoothness
      measure seen   after applying the methods of this paper. 
       All these changes
    are positive, indicating that the resulting landscapes are much smoother.
    For an intuition of what these numbers mean,    the TOMCAT change of 36.35\% results in Figure \ref{fig:loss}  
    changing to Figure \ref{fig:gain}.
    
    Hence we say:

\begin{formal}
\ans{2} Label, parameter, instance and boundary engineering can simplify the
internal structure of   training
data.
\end{formal}

\subsection*{
\rqn{3}  Does  GHOST2's  combination of {\em instance}, {\em label}, {\em boundary}  and {\em parameter}  improve predictive performance?}

\begin{figure}[!t]
   \begin{center}\includegraphics[width=.3\textwidth]{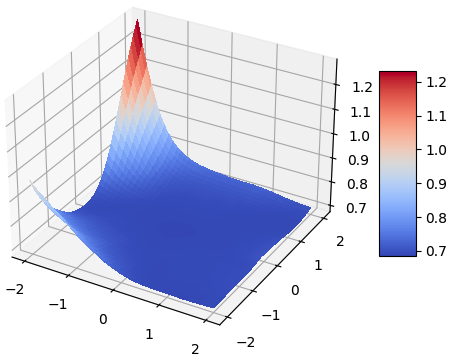} \end{center} 
    \caption{ Error landscape in the TOMCAT  after applying the treatments of \S\ref{rx}. To understand the simplifications achieved via our methods, the reader might find it insightful to  compare this figure against Figure~\ref{fig:loss}. }
    \label{fig:gain}
\end{figure}

 Table~\ref{summary} shows the performance
 improvements   after
 smoothing out our training
 data from (e.g.)
 Figure \ref{fig:loss} to 
 Figure \ref{fig:gain}. On 4/8 datasets, we achieve perfect scores. Moreover, we showed through an ablation study that each of the components of GHOST2 is necessary. For example, row A3 in Table \ref{tab:results} is another piece of evidence that hyper-parameter optimization is necessary. The feedforward networks of our approach outperformed more complex learners (CNNs and CodeBERT)--we refer the reader to rows B2, C2, and CodeBERT in Table \ref{tab:results}. On the other hand, going too simple for traditional learners leads to A7, which suffers from poor precision scores.
 Given those large improvements, we say: 
 
\begin{formal}
\ans{3} 
Detectors of actionable 
 static code warnings  work much better 
 when learned from smoothed
 training data.
 \end{formal}

\subsection*{  
\rqn{4}  Are all parts of GHOST2 necessary; i.e. would
something simpler also achieve the overall goal?}

We presented an ablation study that showed that each part of GHOST2 was necessary. Among the 13 treatments that we tested, GHOST2 was the only one that consistently scored highly in precision, AUC, and recall, while also generally having low false alarm rates. The crux of our ablation study was that each component of GHOST2 works with the others to produce a strong performer.
 
 Based on the above ablation study results, we say:
 \begin{formal}
\ans{4} Ignoring any of
part of
instance,  label,  boundary  or parameter engineering leads to worse results than
using all parts (at least for the purpose of recognizing actionable static code
warnings).
\end{formal}

 \subsection*{
\rqn{5}  Are larger training sets necessary (for the task of recognizing actionable static code warnings)?}
 
 In the above discussion, when we presented
Table~\ref{tab:data}, it was noted that several
of the train/tests used in this study were very small. At that time, we expressed a concern that, possibly, our   data sets explored   were too small for effective learning.

This turned out not to be the case.
Recall that in 
  Table~\ref{tab:results},
  the data set were sorted left-to-right from smallest to largest training set size. There is no pattern there  that smaller data sets perform worse than large ones.  In fact-- quite the opposite: the smaller data sets were always associated with better performance than those seen on right-left-side.  Hence we say:

  \begin{formal}
\ans{5} The methods of this paper are effective,
even for very small data sets.
\end{formal}

This is a surprising result since one of the truisms of data mining is ``the more data, the better'':
Researchers in linear regression have a rule of thumb that every independent attribute implies needing an additional  10 to  20 training examples~\cite{peduzzi1996simulation,austin2017events}. By that reasoning,   our data sets with 260 attributes  would need 2600 to 5200 examples before a learner could achieve competency.

That said, those rules of thumb for regression models were developed where:
\bi
\item
Inputs are naturally occurring (or handcrafted) attributes (not the more nuanced hyper-dimensions found by neural methods);
\item
The response variable (output) is a continuous variable that can vary over a large numeric range (and not the two labels we are exploring). 
\ei
A better conceptual model for our work, that could explain our success when reasoning about small data,   might be ``pin the tail on the donkey'' where there are a handful of green and red pins already in place and we have to add in a ribbon that separates (say) the different colors. Note that since we are using neural, we are free to pull that ribbon across any dimension we like, or even infer a new dimension, if we want.
We would argue that  in this second conceptualization, it is totally reasonable to expect that such a ribbon can be found {\em especially} when we are trying to separate only a handful of pins.

 \section{Threats to Validity}
\label{sec:threats}

 As with any empirical study, biases can affect the final results. Therefore, any conclusions made from this work must be considered with the following issues in mind: 
 
 {\em 1. Sampling bias } threatens any classification experiment; i.e., what matters there may not be true here. For example, the data sets used here comes prior work
 and, possibly, if we explored other data sets we might reach other conclusions.
On the other hand, {\em repeatability} is an important part of science so we argue
that our decision to use the Kang et al. data is appropriate and respectful to both that prior work
and the scientific method.   
 
 {\em 2. Learner bias:} Machine learning is a large and active field and any single study can only use a small subset of the known  algorithms. Our choice of ``local learning'' tools was explained in \S\ref{rx}. That said, it is important to repeat the comments made there
 that our SMOTEing, SMOOTHing, GHOSTing and DODGEing   operators
 are but one set of  choice within a larger framework of possible approaches to instance, label, boundary, and parameter engineering (respectively).
 As SE research matures, we foresee that our framework will become a workbench   within which researchers replace some/all of these treatments with better options. That said, in defence of the current options, we note that our ablation study showed that removing any of them can lead to worse results. 

 \BLUE
 
\respto{2a2.1}
Another threat to validity is the   stochastic nature of SMOOTH (since points are selected at random).
We tested if  SMOOTH leads to conclusion instability as follows. SMOOTH uses a k-d tree to perform hierarchical clustering. We modify it so that it returns the median positions of the leaf clusters. By running this operation 20 times, we have a set of medians. We say that our conclusions are stable if those medians are stable. We conjectured that over each run, because the process is stochastic, that these medians would move slightly; therefore, we ran the following steps:
\begin{itemize}
    \item From the k-d tree clustering, obtain the number of leaf clusters, call this $k$
    \item Run SMOOTH 20 times, and collect a list of the medians. Each set of medians should have $k$ elements, where each element is $d-$dimensional. That is, we have $20\times k$ points, each $d-$dimensional.
    \item On these $20k$ points, we run $k-$means clustering, using $k$ from Step 1 as the number of clusters.
    \item For each cluster, we find the median, and then compute the deviations (via the L1 norm \cite{aggarwal2001surprising}) of each point in that cluster to this median. This gives us a set of deviations, of which we compute the median.
    \item We compute this median as a percentage of the L1-norm of the dataset 
\end{itemize}
Overall, these percentages are extremely low, with the highest being 0.52\%. This suggests that over 20 repeats, while the median does shift, it does so in very small proportions relative to the overall dataset. Therefore, we believe this should not have a meaningful impact on our results.
\BLACK

 {\em 3. Parameter bias}: Learners are controlled by parameters and the resulting performance can change dramatically if those parameters are changed. Accordingly,
 in this paper, our recommended methods (from Table~\ref{rx}) includes parameter engineering methods to find good parameter settings for our different data sets.

 {\em 4. Evaluation bias:} This paper use four evaluation criteria (precision, AUC, false alarm rate, and recall) and it is certainly true that by other measures, our results might not work be seen to work as as well. In defence of our current selection, we note that we use these measures since they let us compare our new results to prior work (who reported their results using the same measures).
 
 Also, to repeat a remark made previously, another evaluation bias was how we separated data into train/test.  In other papers, we have run repeated
trials with multiple 80:20 splits for training:test data. This
was not here since some of our data sets are too small (see
the first few rows of Table 5) that any reduction in the
training set size might disadvantage the learning process.
Hence, our conclusion do
{\bf not} come
from (say) some 5$\times$5 cross-validation experiments.
That experiment would take 20 months of CPU to execute  and generate comparison problems since our test sets would be of radically different sizes (since some of our data sets are tiny)

Hence, the external validity claims of this paper come from
patterns seen in eight different software projects.
 Suppose you wanted
to refute the hypothesis that our recommended treatment  (A1) is better than the rest. To do so, we need to to find examples that satisfied three tests:
\bi
\item
{\em Test1}: the performance metric collected from the   other methods has to be better; 
\item
{\em Test2}: the size of the difference in the performance metric has to be larger than a small effect (this is the effect size test);
\item
{\em Test3}:  the populations from which the performance metrics are drawn are distinguishable (this is the statistical significance test).
\ei
Note that there is no point doing the significance and effect size tests {\em unless} we can first find evidence of better performance.  Hence, it should be noted that {\em Test2} and {\em Test3}  are unnecessary if {\em Test1} fails. Accordingly, we now turn our attention to {\em Test1}  and the results from   Table~\ref{tab:results}.

In  Table~\ref{tab:results}., suppose to define ``better''
we look at the 13 rows in the four tables (precision, AUC, false alarm and recall tables).
Suppose further we  define ``better'' as follows:
\bi
\item
Row X is ``better than A1''  if {\em in $N$ cells of row X, the performance metric is better than A1. }
\ei
(Aside: recall that ``better'' is different for different measures; for false alarm, ``better'' means ``less'' while for the others ``better'' means ``more''.)

At first it might be tempting to use $N>4$  (since we are dealing with eight data sets and $N>4$ would mean we are saying ``in the majority case'').
It turns out that  this $N>4$ definition of ``better'' is almost unachievable since,
reading
Table~\ref{tab:results} we see that:
\begin{itemize}
\item
There is only one  example
of  better $N\ge 4$ rows (for C1 (DODGE) AUC);
\end{itemize}
That said there are 
 six examples
where $N=3$ cells are better
in some row X than A1\footnote{One: A6 for AUC;
two: C1 for AUC;
three: A6 for false alarm rate;
four: A7 for false alarm rate;
five: C1 for false alarm rate; 
six: A5 for recall.}.
That is to say, if  we define $N>=3$ as our threshold for  row X is ``better than A12'' then  in only 7/48 treatments would it make sense to apply a significance and effect size test. 

In summary:   at the $N>=3$ level,  in 41/48 of our experiments, we cannot refute the hypothesis that A12 is better than other treatments. 

\section{Discussion}
\label{sec:discussion}

This discussion section steps back from the above to make some more general points.

We suggest that this paper should lead to a new way of training
newcomers in software analytics:
\bi
\item
Our results show that there is much value in decades-old learning technology
(feedforward networks). Hence, we say that when we train newcomers to the field of software analytics,
we should certainly train them in the latest techniques (deep learning, CodeBERT, etc).
\item
That said,
we should also ensure that they know of prior work since (as shown above), sometimes
those older methods still have currency. 
For example,
if some learner is faster to run, then it is easier to tune. 
Hence, as shown above, it can be possible for old techniques to do better than new ones, just by tuning.  
\ei
   For future work, it would be useful to check what other SE domains
    simpler, faster, learners  (plus some tuning)
   out-perform more complex learning methods.

That said, we offer the following cautionary note about tuning.
Hyper-parameter  optimization (HPO, which  we have call ``parameter engineering'' in this paper)
  has received much recent attention in the SE literature~\cite{agrawal2019dodge,yedida2021value,agrawal2021simpler} 
We have shown here that reliance on {\em just} HPO can be foolhardy
since better results can be obtained by the 
judicious  use of HPO combined with more nuanced approaches that actually reflect the particulars of the
current problem (e.g. our treatments that adjusted different parts of the data in different ways).
As to how much to study the internals of a learner,
we showed above  that there are many choices deep within a learner than can greatly improve predictive performance.  Hence we say that it   is very important to know the internals of a learner and how to adjust them.
In our opinion, all too often, software engineers use AI tools as ``black boxes'' with little
understanding of their internal structure. 

Our results also  doubt some of the truisms of our field. 
For example:
\bi
\item There is much recent work on big data research in SE, the premise being that ``the more data, the better''. We certainly do not dispute that but
our  results do show that it is possible to achieve good results with very small data sets.
\item
There is much work in software analytics suggesting that  deep learning is a superior method for analyzing data \cite{yedida2021value, wang2018deep, li2017cclearner, white2015deep}. Yet when we tried that here, we found that a decades-old neural net architecture (feed-forward networks, discussed in Table \ref{tab:nnhere}) significantly out-performed deep learners. 
\ei
For newcomers to the field of software analytics,
truisms   might be useful. But better results
might be obtained when teams of data scientists
combine to suggest  multiple techniques -- some of which ignore supposedly tried-and-true truisms.

\section{Conclusion}
\label{sec:conclusion}

Static analysis tools often
suffer from a large number of false alarms that are deemed
to be unactionable~\cite{tomassi2021real}. Hence, developers often ignore
many of their warnings. Prior work by 
Yang et al.~\cite{yang2021learning} attempted to build predictors for
actionable warnings but,
as shown by Kang et al.~\cite{kang2022detecting}, that study used
poorly labelled data. 

This paper extends the  Kang et al. result as follows.
 Table~\ref{tab:initial_svm} shows that building
 models for this domain is a challenging task. 
 The discussion section of \S\ref{problem}
 conjectured that for the purposes of detecting
 actionable static code warnings, standard
 data miners can not handle the complexities of
 the   decision boundary. More specifically, 
 we argued that:
 \begin{quote}
 {\em 
 For complex data, \underline{\bf global} treatments  
   perform worse  than \underline{\bf localized} treatments
 which   adjust different parts of the  landscape in 
 different ways.}
 \end{quote}
 \S\ref{rx} proposed four such localized treatments, 
 which we called instance, parameter, label and boundary engineering.
 
 These treatments were tested on the data generated by Kang et al.
 (which in turn, was generated by fixing the prior missteps of Yang et al.).
 On experimentation, it was shown that the combination of
 all our treatments (in the ``A1'' results of 
  Table~\ref{tab:treatments}) performed much better than than the prior
  results seen in  Table~\ref{tab:initial_svm}. 
 As to why these treatments before so well, the analysis
 of Table~\ref{tab:fuzzy1} showed that instance, parameter, label and boundary engineering
 did in fact remove complex shapes in our decision boundaries.
  As to the relative merits of instance versus parameter versus label versus boundary engineering,
an ablation study showed that using all these treatments produces better predictions that 
   alternative treatments that ignored any part.

Finally, we comment here on the value of different teams working together.
The specific result reported in this paper
is about 
how to recognize and avoid
static code analysis false alarms. That said,
there is a more general takeaway.
Science is meant to be about a community critiquing and improving each other's ideas. Here, we offer a successful example of such a community interaction where teams from Singapore and the US successfully worked together. Initially, in a 2022 paper~\cite{kang2022detecting}, the Singapore team identified issues with the data that result in substantially lower performance of the previously-reported best predictor of actionable warnings~\cite{wang2018there, yang2021learning,yang2021understanding}. Subsequently, in this paper,
both teams combined  to produce new results that clarified and improved the old work.
That teamwork leads us to trying    methods which, according to the truisms of our field, should not have worked.   The teamwork that generated this paper
  should be routine, and not some rare exceptional case.

\section*{Acknowledgments}
This work was partially supported by an NSF Grant \#1908762.

{\footnotesize \bibliographystyle{plainnat}
 
\bibliography{tim.bib,rahul_rahul.bib,hongjin_hongjin.bib,rahul_other.bib}

\begin{thebibliography}{75}
\providecommand{\natexlab}[1]{#1}
\providecommand{\url}[1]{\texttt{#1}}
\expandafter\ifx\csname urlstyle\endcsname\relax
  \providecommand{\doi}[1]{doi: #1}\else
  \providecommand{\doi}{doi: \begingroup \urlstyle{rm}\Url}\fi

\bibitem[Aggarwal et~al.(2001)Aggarwal, Hinneburg, and
  Keim]{aggarwal2001surprising}
Charu~C Aggarwal, Alexander Hinneburg, and Daniel~A Keim.
\newblock On the surprising behavior of distance metrics in high dimensional
  space.
\newblock In \emph{International conference on database theory}, pages
  420--434. Springer, 2001.

\bibitem[Agrawal and Menzies(2018)]{agrawal2018better}
Amritanshu Agrawal and Tim Menzies.
\newblock Is" better data" better than" better data miners"?
\newblock In \emph{2018 IEEE/ACM 40th International Conference on Software
  Engineering (ICSE)}, pages 1050--1061. IEEE, 2018.

\bibitem[Agrawal et~al.(2019)Agrawal, Fu, Chen, Shen, and
  Menzies]{agrawal2019dodge}
Amritanshu Agrawal, Wei Fu, Di~Chen, Xipeng Shen, and Tim Menzies.
\newblock How to “dodge” complex software analytics.
\newblock \emph{IEEE Transactions on Software Engineering}, 47\penalty0
  (10):\penalty0 2182--2194, 2019.

\bibitem[Agrawal et~al.(2021)Agrawal, Yang, Agrawal, Yedida, Shen, and
  Menzies]{agrawal2021simpler}
Amritanshu Agrawal, Xueqi Yang, Rishabh Agrawal, Rahul Yedida, Xipeng Shen, and
  Tim Menzies.
\newblock Simpler hyperparameter optimization for software analytics: why, how,
  when.
\newblock \emph{IEEE Transactions on Software Engineering}, 2021.

\bibitem[Austin and Steyerberg(2017)]{austin2017events}
Peter~C Austin and Ewout~W Steyerberg.
\newblock Events per variable (epv) and the relative performance of different
  strategies for estimating the out-of-sample validity of logistic regression
  models.
\newblock \emph{Statistical methods in medical research}, 26\penalty0
  (2):\penalty0 796--808, 2017.

\bibitem[Ayewah and Pugh(2010)]{ayewah2010google}
Nathaniel Ayewah and William Pugh.
\newblock The google findbugs fixit.
\newblock In \emph{Proceedings of the 19th International Symposium on Software
  Testing and Analysis}, pages 241--252, 2010.

\bibitem[Banerjee et~al.(2019)Banerjee, Clapp, and
  Sridharan]{banerjee2019nullaway}
Subarno Banerjee, Lazaro Clapp, and Manu Sridharan.
\newblock Nullaway: Practical type-based null safety for java.
\newblock In \emph{Proceedings of the 2019 27th ACM Joint Meeting on European
  Software Engineering Conference and Symposium on the Foundations of Software
  Engineering}, pages 740--750, 2019.

\bibitem[Barkan et~al.(2021)Barkan, Hazan, and Ratner]{barkan2021reduce}
Ella Barkan, Alon Hazan, and Vadim Ratner.
\newblock Reduce discrepancy of human annotators in medical imaging by
  automatic visual comparison to similar cases, February~9 2021.
\newblock US Patent 10,916,343.

\bibitem[Beller et~al.(2016)Beller, Bholanath, McIntosh, and
  Zaidman]{beller2016analyzing}
Moritz Beller, Radjino Bholanath, Shane McIntosh, and Andy Zaidman.
\newblock Analyzing the state of static analysis: A large-scale evaluation in
  open source software.
\newblock In \emph{2016 IEEE 23rd International Conference on Software
  Analysis, Evolution, and Reengineering (SANER)}, volume~1, pages 470--481.
  IEEE, 2016.

\bibitem[Bergstra and Bengio(2012)]{Bergstra12}
James Bergstra and Yoshua Bengio.
\newblock Random search for hyper-parameter optimization.
\newblock \emph{J. Mach. Learn. Res.}, 13\penalty0 (null):\penalty0 281–305,
  feb 2012.
\newblock ISSN 1532-4435.

\bibitem[Berthelot et~al.(2019)Berthelot, Carlini, Goodfellow, Papernot,
  Oliver, and Raffel]{berthelot2019mixmatch}
David Berthelot, Nicholas Carlini, Ian Goodfellow, Nicolas Papernot, Avital
  Oliver, and Colin~A Raffel.
\newblock Mixmatch: A holistic approach to semi-supervised learning.
\newblock \emph{Advances in Neural Information Processing Systems}, 32, 2019.

\bibitem[Calcagno et~al.(2015)Calcagno, Distefano, Dubreil, Gabi, Hooimeijer,
  Luca, O’Hearn, Papakonstantinou, Purbrick, and
  Rodriguez]{calcagno2015moving}
Cristiano Calcagno, Dino Distefano, J{\'e}r{\'e}my Dubreil, Dominik Gabi,
  Pieter Hooimeijer, Martino Luca, Peter O’Hearn, Irene Papakonstantinou, Jim
  Purbrick, and Dulma Rodriguez.
\newblock Moving fast with software verification.
\newblock In \emph{NASA Formal Methods Symposium}, pages 3--11. Springer, 2015.

\bibitem[Chakraborty et~al.(2022)Chakraborty, Majumder, and Tu]{fairssl}
J.~Chakraborty, S.~Majumder, and H.~Tu.
\newblock Can we achieve fairness using semi-supervised learning?
\newblock \emph{Fairware}, 2022.

\bibitem[Chawla et~al.(2002)Chawla, Bowyer, Hall, and
  Kegelmeyer]{chawla2002smote}
Nitesh~V Chawla, Kevin~W Bowyer, Lawrence~O Hall, and W~Philip Kegelmeyer.
\newblock Smote: synthetic minority over-sampling technique.
\newblock \emph{Journal of artificial intelligence research}, 16:\penalty0
  321--357, 2002.

\bibitem[Christakis and Bird(2016)]{ChristakisB16}
Maria Christakis and Christian Bird.
\newblock What developers want and need from program analysis: an empirical
  study.
\newblock In \emph{Proceedings of the 31st {IEEE/ACM} International Conference
  on Automated Software Engineering, {ASE} 2016, Singapore, September 3-7,
  2016}, pages 332--343. {ACM}, 2016.

\bibitem[Cohen(1995)]{cohen1995empirical}
Paul~R Cohen.
\newblock \emph{Empirical methods for artificial intelligence}, volume 139.
\newblock MIT press Cambridge, 1995.

\bibitem[Cordeiro and Carneiro(2020)]{cordeiro2020survey}
Filipe~R Cordeiro and Gustavo Carneiro.
\newblock A survey on deep learning with noisy labels: How to train your model
  when you cannot trust on the annotations?
\newblock In \emph{2020 33rd SIBGRAPI conference on graphics, patterns and
  images (SIBGRAPI)}, pages 9--16. IEEE, 2020.

\bibitem[Croft et~al.(2021)Croft, Newlands, Chen, and
  Babar]{croft2021empirical}
Roland Croft, Dominic Newlands, Ziyu Chen, and M~Ali Babar.
\newblock An empirical study of rule-based and learning-based approaches for
  static application security testing.
\newblock In \emph{Proceedings of the 15th ACM/IEEE International Symposium on
  Empirical Software Engineering and Measurement (ESEM)}, pages 1--12, 2021.

\bibitem[Deb et~al.(2005)Deb, Mohan, and Mishra]{Deb05}
Kalyan Deb, Manikanth Mohan, and Shikhar Mishra.
\newblock Evaluating the $\epsilon$-dominance based multi-objective
  evolutionary algorithm for a quick computation of pareto-optimal solutions.
\newblock \emph{Evolutionary computation}, 13:\penalty0 501--25, 02 2005.
\newblock \doi{10.1162/106365605774666895}.

\bibitem[Feng et~al.(2020)Feng, Guo, Tang, Duan, Feng, Gong, Shou, Qin, Liu,
  Jiang, et~al.]{feng2020codebert}
Zhangyin Feng, Daya Guo, Duyu Tang, Nan Duan, Xiaocheng Feng, Ming Gong, Linjun
  Shou, Bing Qin, Ting Liu, Daxin Jiang, et~al.
\newblock Codebert: A pre-trained model for programming and natural languages.
\newblock \emph{arXiv preprint arXiv:2002.08155}, 2020.

\bibitem[Habib and Pradel(2018)]{habib2018many}
Andrew Habib and Michael Pradel.
\newblock How many of all bugs do we find? a study of static bug detectors.
\newblock In \emph{2018 33rd IEEE/ACM International Conference on Automated
  Software Engineering (ASE)}, pages 317--328. IEEE, 2018.

\bibitem[Hanam et~al.(2014)Hanam, Tan, Holmes, and Lam]{hanam2014finding}
Quinn Hanam, Lin Tan, Reid Holmes, and Patrick Lam.
\newblock Finding patterns in static analysis alerts: improving actionable
  alert ranking.
\newblock In \emph{Proceedings of the 11th working conference on mining
  software repositories}, pages 152--161, 2014.

\bibitem[Heckman and Williams(2008)]{heckman2008establishing}
Sarah Heckman and Laurie Williams.
\newblock On establishing a benchmark for evaluating static analysis alert
  prioritization and classification techniques.
\newblock In \emph{Proceedings of the Second ACM-IEEE international symposium
  on Empirical software engineering and measurement}, pages 41--50, 2008.

\bibitem[Heckman and Williams(2009)]{heckman2009model}
Sarah Heckman and Laurie Williams.
\newblock A model building process for identifying actionable static analysis
  alerts.
\newblock In \emph{2009 International conference on software testing
  verification and validation}, pages 161--170. IEEE, 2009.

\bibitem[Imtiaz et~al.(2019)Imtiaz, Murphy, and Williams]{imtiaz2019developers}
Nasif Imtiaz, Brendan Murphy, and Laurie Williams.
\newblock How do developers act on static analysis alerts? an empirical study
  of coverity usage.
\newblock In \emph{2019 IEEE 30th International Symposium on Software
  Reliability Engineering (ISSRE)}, pages 323--333. IEEE, 2019.

\bibitem[Ioffe and Szegedy(2015)]{ioffe2015batch}
Sergey Ioffe and Christian Szegedy.
\newblock Batch normalization: Accelerating deep network training by reducing
  internal covariate shift.
\newblock In \emph{International conference on machine learning}, pages
  448--456. PMLR, 2015.

\bibitem[Johnson et~al.(2013)Johnson, Song, Murphy-Hill, and
  Bowdidge]{johnson2013don}
Brittany Johnson, Yoonki Song, Emerson Murphy-Hill, and Robert Bowdidge.
\newblock Why don't software developers use static analysis tools to find bugs?
\newblock In \emph{2013 35th International Conference on Software Engineering
  (ICSE)}, pages 672--681. IEEE, 2013.

\bibitem[Kallingal~Joshy et~al.(2021)Kallingal~Joshy, Chen, Steenhoek, and
  Le]{kallingal2021validating}
Ashwin Kallingal~Joshy, Xueyuan Chen, Benjamin Steenhoek, and Wei Le.
\newblock Validating static warnings via testing code fragments.
\newblock In \emph{Proceedings of the 30th ACM SIGSOFT International Symposium
  on Software Testing and Analysis}, pages 540--552, 2021.

\bibitem[Kang et~al.(2022)Kang, Aw, and Lo]{kang2022detecting}
Hong~Jin Kang, Khai~Loong Aw, and David Lo.
\newblock Detecting false alarms from automatic static analysis tools: How far
  are we?
\newblock \emph{arXiv preprint arXiv:2202.05982}, 2022.

\bibitem[Kharkar et~al.(2022)Kharkar, Moghaddam, Jin, Liu, Shi, Clement, and
  Sundaresan]{kharkar2022learning}
Anant Kharkar, Roshanak~Zilouchian Moghaddam, Matthew Jin, Xiaoyu Liu, Xin Shi,
  Colin Clement, and Neel Sundaresan.
\newblock Learning to reduce false positives in analytic bug detectors.
\newblock In \emph{Proceedings of the IEEE/ACM International Conference on
  Software Engineering 2022}, 2022.

\bibitem[Kim et~al.(2022)Kim, Raghothaman, and Heo]{kim2022learning}
Hyunsu Kim, Mukund Raghothaman, and Kihong Heo.
\newblock Learning probabilistic models for static analysis alarms.
\newblock In \emph{Proceedings of the IEEE/ACM International Conference on
  Software Engineering 2022}, 2022.

\bibitem[Kim and Ernst(2007{\natexlab{a}})]{kim07}
Sunghun Kim and Michael~D. Ernst.
\newblock Which warnings should i fix first?
\newblock In \emph{Proceedings of the the 6th Joint Meeting of the European
  Software Engineering Conference and the ACM SIGSOFT Symposium on The
  Foundations of Software Engineering}, ESEC-FSE '07, page 45–54, New York,
  NY, USA, 2007{\natexlab{a}}. Association for Computing Machinery.
\newblock ISBN 9781595938114.
\newblock \doi{10.1145/1287624.1287633}.
\newblock URL \url{https://doi.org/10.1145/1287624.1287633}.

\bibitem[Kim and Ernst(2007{\natexlab{b}})]{kim2007warnings}
Sunghun Kim and Michael~D Ernst.
\newblock Which warnings should {I} fix first?
\newblock In \emph{Proceedings of the the 6th joint meeting of the European
  Software Engineering Conference and the ACM SIGSOFT Symposium on The
  Foundations of Software Engineering}, pages 45--54, 2007{\natexlab{b}}.

\bibitem[Kingma et~al.(2014)Kingma, Mohamed, Jimenez~Rezende, and
  Welling]{kingma2014semi}
Durk~P Kingma, Shakir Mohamed, Danilo Jimenez~Rezende, and Max Welling.
\newblock Semi-supervised learning with deep generative models.
\newblock \emph{Advances in neural information processing systems}, 27, 2014.

\bibitem[Krizhevsky et~al.(2012)Krizhevsky, Sutskever, and
  Hinton]{krizhevsky2012imagenet}
Alex Krizhevsky, Ilya Sutskever, and Geoffrey~E Hinton.
\newblock Imagenet classification with deep convolutional neural networks.
\newblock \emph{Advances in neural information processing systems}, 25, 2012.

\bibitem[Kumar()]{kumar20}
Ajitesh Kumar.
\newblock Svm rbf kernel parameters with code examples.
\newblock Available on-line at
  \url{https://dzone.com/articles/using-jsonb-in-postgresql-how-to-effectively-store-1}.

\bibitem[LeCun et~al.(1989)LeCun, Boser, Denker, Henderson, Howard, Hubbard,
  and Jackel]{lecun1989backpropagation}
Yann LeCun, Bernhard Boser, John~S Denker, Donnie Henderson, Richard~E Howard,
  Wayne Hubbard, and Lawrence~D Jackel.
\newblock Backpropagation applied to handwritten zip code recognition.
\newblock \emph{Neural computation}, 1\penalty0 (4):\penalty0 541--551, 1989.

\bibitem[Li et~al.(2018)Li, Xu, Taylor, Studer, and
  Goldstein]{li2018visualizing}
Hao Li, Zheng Xu, Gavin Taylor, Christoph Studer, and Tom Goldstein.
\newblock Visualizing the loss landscape of neural nets.
\newblock \emph{Advances in Neural Information Processing Systems}, 31, 2018.

\bibitem[Li et~al.(2017)Li, Feng, Zhuang, Meng, and Ryder]{li2017cclearner}
Liuqing Li, He~Feng, Wenjie Zhuang, Na~Meng, and Barbara Ryder.
\newblock Cclearner: A deep learning-based clone detection approach.
\newblock In \emph{2017 IEEE International Conference on Software Maintenance
  and Evolution (ICSME)}, pages 249--260. IEEE, 2017.

\bibitem[Liang et~al.(2010)Liang, Wu, Wu, Wang, Xie, and
  Mei]{liang2010automatic}
Guangtai Liang, Ling Wu, Qian Wu, Qianxiang Wang, Tao Xie, and Hong Mei.
\newblock Automatic construction of an effective training set for prioritizing
  static analysis warnings.
\newblock In \emph{Proceedings of the IEEE/ACM international conference on
  Automated software engineering}, pages 93--102, 2010.

\bibitem[Ma et~al.(2019)Ma, Liu, Fang, and Simoncelli]{ma2019blind}
Kede Ma, Xuelin Liu, Yuming Fang, and Eero~P Simoncelli.
\newblock Blind image quality assessment by learning from multiple annotators.
\newblock In \emph{2019 IEEE International Conference on Image Processing
  (ICIP)}, pages 2344--2348. IEEE, 2019.

\bibitem[McNicol(2005)]{mcnicol2005primer}
Don McNicol.
\newblock \emph{A primer of signal detection theory}.
\newblock Psychology Press, 2005.

\bibitem[Montufar et~al.(2014)Montufar, Pascanu, Cho, and
  Bengio]{montufar2014number}
Guido~F Montufar, Razvan Pascanu, Kyunghyun Cho, and Yoshua Bengio.
\newblock On the number of linear regions of deep neural networks.
\newblock \emph{Advances in Neural Information Processing Systems}, 27, 2014.

\bibitem[Muske and Serebrenik(2020)]{muske2020techniques}
Tukaram Muske and Alexander Serebrenik.
\newblock Techniques for efficient automated elimination of false positives.
\newblock In \emph{2020 IEEE 20th International Working Conference on Source
  Code Analysis and Manipulation (SCAM)}, pages 259--263. IEEE, 2020.

\bibitem[Nachtigall et~al.(2022)Nachtigall, Schlichtig, and
  Bodden]{nachtigall2022large}
Marcus Nachtigall, Michael Schlichtig, and Eric Bodden.
\newblock A large-scale study of usability criteria addressed by static
  analysis tools.
\newblock In \emph{Proceedings of the 31st ACM SIGSOFT International Symposium
  on Software Testing and Analysis}, pages 532--543, 2022.

\bibitem[Nam et~al.(2019)Nam, Wang, Xi, and Tan]{nam2019bug}
Jaechang Nam, Song Wang, Yuan Xi, and Lin Tan.
\newblock A bug finder refined by a large set of open-source projects.
\newblock \emph{Information and Software Technology}, 112:\penalty0 164--175,
  2019.

\bibitem[Panichella et~al.(2015)Panichella, Arnaoudova, Di~Penta, and
  Antoniol]{panichella2015would}
Sebastiano Panichella, Venera Arnaoudova, Massimiliano Di~Penta, and Giuliano
  Antoniol.
\newblock Would static analysis tools help developers with code reviews?
\newblock In \emph{2015 IEEE 22nd International Conference on Software
  Analysis, Evolution, and Reengineering (SANER)}, pages 161--170. IEEE, 2015.

\bibitem[Peduzzi et~al.(1996)Peduzzi, Concato, Kemper, Holford, and
  Feinstein]{peduzzi1996simulation}
Peter Peduzzi, John Concato, Elizabeth Kemper, Theodore~R Holford, and Alvan~R
  Feinstein.
\newblock A simulation study of the number of events per variable in logistic
  regression analysis.
\newblock \emph{Journal of clinical epidemiology}, 49\penalty0 (12):\penalty0
  1373--1379, 1996.

\bibitem[Ribeiro et~al.(2016)Ribeiro, Singh, and Guestrin]{ribeiro2016should}
Marco~Tulio Ribeiro, Sameer Singh, and Carlos Guestrin.
\newblock " why should i trust you?" explaining the predictions of any
  classifier.
\newblock In \emph{Proceedings of the 22nd ACM SIGKDD international conference
  on knowledge discovery and data mining}, pages 1135--1144, 2016.

\bibitem[Rumelhart et~al.(1986)Rumelhart, Hinton, and
  Williams]{rumelhart1986learning}
David~E Rumelhart, Geoffrey~E Hinton, and Ronald~J Williams.
\newblock Learning representations by back-propagating errors.
\newblock \emph{nature}, 323\penalty0 (6088):\penalty0 533--536, 1986.

\bibitem[Ruthruff et~al.(2008)Ruthruff, Penix, Morgenthaler, Elbaum, and
  Rothermel]{ruthruff2008predicting}
Joseph Ruthruff, John Penix, J~Morgenthaler, Sebastian Elbaum, and Gregg
  Rothermel.
\newblock Predicting accurate and actionable static analysis warnings.
\newblock In \emph{2008 ACM/IEEE 30th International Conference on Software
  Engineering}, pages 341--350. IEEE, 2008.

\bibitem[Sadowski et~al.(2018)Sadowski, Aftandilian, Eagle, Miller-Cushon, and
  Jaspan]{sadowski2018lessons}
Caitlin Sadowski, Edward Aftandilian, Alex Eagle, Liam Miller-Cushon, and Ciera
  Jaspan.
\newblock Lessons from building static analysis tools at google.
\newblock \emph{Communications of the ACM}, 61\penalty0 (4):\penalty0 58--66,
  2018.

\bibitem[Santurkar et~al.(2018)Santurkar, Tsipras, Ilyas, and
  Madry]{santurkar2018does}
Shibani Santurkar, Dimitris Tsipras, Andrew Ilyas, and Aleksander Madry.
\newblock How does batch normalization help optimization?
\newblock \emph{Advances in Neural Information Processing Systems}, 31, 2018.

\bibitem[Srivastava et~al.(2014)Srivastava, Hinton, Krizhevsky, Sutskever, and
  Salakhutdinov]{srivastava2014dropout}
Nitish Srivastava, Geoffrey Hinton, Alex Krizhevsky, Ilya Sutskever, and Ruslan
  Salakhutdinov.
\newblock Dropout: a simple way to prevent neural networks from overfitting.
\newblock \emph{The journal of machine learning research}, 15\penalty0
  (1):\penalty0 1929--1958, 2014.

\bibitem[Thung et~al.(2012)Thung, Lo, Jiang, Rahman, Devanbu,
  et~al.]{thung2012extent}
Ferdian Thung, David Lo, Lingxiao Jiang, Foyzur Rahman, Premkumar~T Devanbu,
  et~al.
\newblock To what extent could we detect field defects? an empirical study of
  false negatives in static bug finding tools.
\newblock In \emph{2012 Proceedings of the 27th IEEE/ACM International
  Conference on Automated Software Engineering}, pages 50--59. IEEE, 2012.

\bibitem[Tiganov et~al.(2022)Tiganov, Do, and Ali]{tiganov2022designing}
Daniil Tiganov, Lisa Nguyen~Quang Do, and Karim Ali.
\newblock Designing uis for static-analysis tools.
\newblock \emph{Communications of the ACM}, 65\penalty0 (2):\penalty0 52--58,
  2022.

\bibitem[Tomassi and Rubio-Gonz{\'a}lez(2021)]{tomassi2021real}
David~A Tomassi and Cindy Rubio-Gonz{\'a}lez.
\newblock On the real-world effectiveness of static bug detectors at finding
  null pointer exceptions.
\newblock In \emph{2021 36th IEEE/ACM International Conference on Automated
  Software Engineering (ASE)}, pages 292--303. IEEE, 2021.

\bibitem[Tu and Menzies(2021)]{frugal}
H.~Tu and T.~Menzies.
\newblock {FRUGAL:} unlocking {SSL} for software analytics.
\newblock \emph{ASE}, 2021.

\bibitem[Tu and Menzies(2022)]{debtfree}
H.~Tu and T.~Menzies.
\newblock {DebtFree:} minimizing labeling cost in self-admitted technical debt
  identification using semi-supervised learning.
\newblock \emph{EMSE}, 2022.

\bibitem[Tu et~al.(2021)Tu, Papadimitriou, Kiran, Wang, Mandal, Deelman, and
  Menzies]{dodge_comparison}
H.~Tu, G.~Papadimitriou, M.~Kiran, C.~Wang, A.~Mandal, E.~Deelman, and
  T.~Menzies.
\newblock Mining workflows for anomalous data transfers.
\newblock \emph{MSR}, 2021.

\bibitem[Tu et~al.(2022)Tu, Yu, and Menzies]{9064604}
H.~Tu, Z.~Yu, and T.~Menzies.
\newblock Better data labelling with emblem (and how that impacts defect
  prediction).
\newblock \emph{IEEE TSE}, 2022.

\bibitem[Vassallo et~al.(2020)Vassallo, Panichella, Palomba, Proksch, Gall, and
  Zaidman]{vassallo2020developers}
Carmine Vassallo, Sebastiano Panichella, Fabio Palomba, Sebastian Proksch,
  Harald~C Gall, and Andy Zaidman.
\newblock How developers engage with static analysis tools in different
  contexts.
\newblock \emph{Empirical Software Engineering}, 25\penalty0 (2):\penalty0
  1419--1457, 2020.

\bibitem[Vaswani et~al.(2017)Vaswani, Shazeer, Parmar, Uszkoreit, Jones, Gomez,
  Kaiser, and Polosukhin]{vaswani2017attention}
Ashish Vaswani, Noam Shazeer, Niki Parmar, Jakob Uszkoreit, Llion Jones,
  Aidan~N Gomez, {\L}ukasz Kaiser, and Illia Polosukhin.
\newblock Attention is all you need.
\newblock \emph{Advances in neural information processing systems}, 30, 2017.

\bibitem[Wang et~al.(2018{\natexlab{a}})Wang, Wang, and Wang]{wang2018there}
Junjie Wang, Song Wang, and Qing Wang.
\newblock Is there a" golden" feature set for static warning identification? an
  experimental evaluation.
\newblock In \emph{Proceedings of the 12th ACM/IEEE international symposium on
  empirical software engineering and measurement}, pages 1--10,
  2018{\natexlab{a}}.

\bibitem[Wang et~al.(2018{\natexlab{b}})Wang, Liu, Nam, and Tan]{wang2018deep}
Song Wang, Taiyue Liu, Jaechang Nam, and Lin Tan.
\newblock Deep semantic feature learning for software defect prediction.
\newblock \emph{IEEE Transactions on Software Engineering}, 46\penalty0
  (12):\penalty0 1267--1293, 2018{\natexlab{b}}.

\bibitem[White(2015)]{white2015deep}
Martin White.
\newblock Deep representations for software engineering.
\newblock In \emph{2015 IEEE/ACM 37th IEEE International Conference on Software
  Engineering}, volume~2, pages 781--783. IEEE, 2015.

\bibitem[Witten et~al.(2011)Witten, Frank, and Hall]{WittenFH11}
Ian~H. Witten, Eibe Frank, and Mark~A. Hall.
\newblock \emph{Data mining: practical machine learning tools and techniques,
  3rd Edition}.
\newblock Morgan Kaufmann, Elsevier, 2011.
\newblock ISBN 9780123748560.
\newblock URL \url{https://www.worldcat.org/oclc/262433473}.

\bibitem[Yang et~al.(2021{\natexlab{a}})Yang, Chen, Yedida, Yu, and
  Menzies]{yang2021learning}
Xueqi Yang, Jianfeng Chen, Rahul Yedida, Zhe Yu, and Tim Menzies.
\newblock Learning to recognize actionable static code warnings (is
  intrinsically easy).
\newblock \emph{Empirical Software Engineering}, 26\penalty0 (3):\penalty0
  1--24, 2021{\natexlab{a}}.

\bibitem[Yang et~al.(2021{\natexlab{b}})Yang, Yu, Wang, and
  Menzies]{yang2021understanding}
Xueqi Yang, Zhe Yu, Junjie Wang, and Tim Menzies.
\newblock Understanding static code warnings: An incremental ai approach.
\newblock \emph{Expert Systems with Applications}, 167:\penalty0 114134,
  2021{\natexlab{b}}.

\bibitem[Yedida and Menzies(2021)]{yedida2021value}
Rahul Yedida and Tim Menzies.
\newblock On the value of oversampling for deep learning in software defect
  prediction.
\newblock \emph{IEEE Transactions on Software Engineering}, 2021.

\bibitem[Yu et~al.(2020)Yu, Fahid, Tu, and Menzies]{jitterbug}
Zhe Yu, Fahmid~Morshed Fahid, Huy Tu, and Tim Menzies.
\newblock Identifying self-admitted technical debts with jitterbug: A two-step
  approach.
\newblock \emph{IEEE Transactions on Software Engineering}, 2020.

\bibitem[Zampetti et~al.(2017)Zampetti, Scalabrino, Oliveto, Canfora, and
  Di~Penta]{zampetti2017open}
Fiorella Zampetti, Simone Scalabrino, Rocco Oliveto, Gerardo Canfora, and
  Massimiliano Di~Penta.
\newblock How open source projects use static code analysis tools in continuous
  integration pipelines.
\newblock In \emph{2017 IEEE/ACM 14th International Conference on Mining
  Software Repositories (MSR)}, pages 334--344. IEEE, 2017.

\bibitem[Zhai et~al.(2019)Zhai, Oliver, Kolesnikov, and Beyer]{zhai2019s4l}
Xiaohua Zhai, Avital Oliver, Alexander Kolesnikov, and Lucas Beyer.
\newblock S4l: Self-supervised semi-supervised learning.
\newblock In \emph{Proceedings of the IEEE/CVF International Conference on
  Computer Vision}, pages 1476--1485, 2019.

\bibitem[Zheng et~al.(2006)Zheng, Williams, Nagappan, Snipes, Hudepohl, and
  Vouk]{zheng2006value}
Jiang Zheng, Laurie Williams, Nachiappan Nagappan, Will Snipes, John~P
  Hudepohl, and Mladen~A Vouk.
\newblock On the value of static analysis for fault detection in software.
\newblock \emph{IEEE Transactions on Software Engineering}, 32\penalty0
  (4):\penalty0 240--253, 2006.

\bibitem[Zhu(2005)]{zhu2005semi}
Xiaojin~Jerry Zhu.
\newblock Semi-supervised learning literature survey.
\newblock 2005.

\end{thebibliography}
}

\begin{IEEEbiography}[{\includegraphics[width=.9in,clip,keepaspectratio]{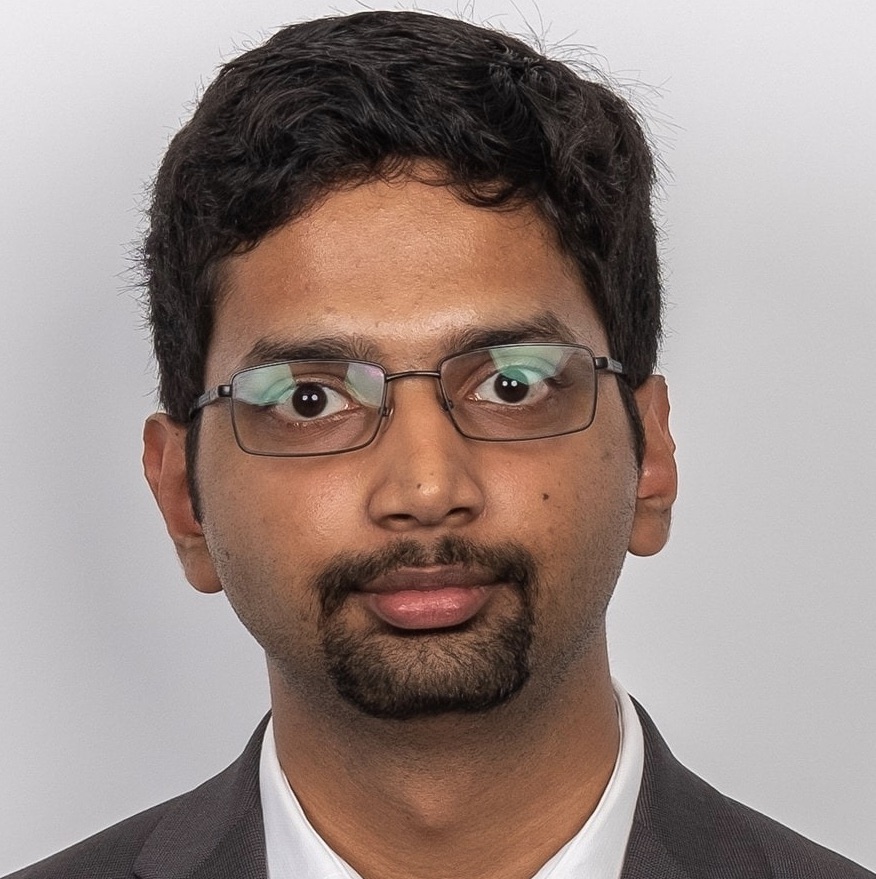}}] {Rahul Yedida} is a PhD student in Computer Science at NC State University. His research interests include automated software engineering and machine learning for software engineering.   \url{https://ryedida.me}.
\end{IEEEbiography}
\vspace{-20mm}
\begin{IEEEbiography}[{\includegraphics[width=.9in,clip,keepaspectratio]{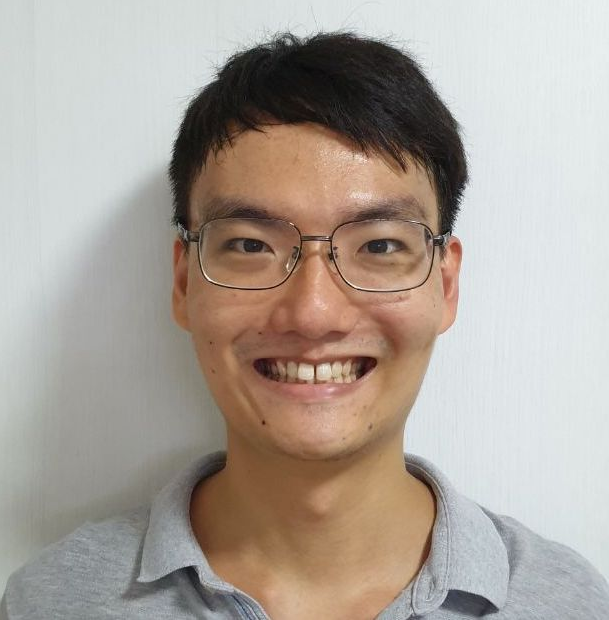}}]{Hong Jin Kang  } is a Ph.D. student at Singapore Management University. His research interests include machine learning for software engineering, and mining rules and specifications.
 \url{https://kanghj.github.io/}.
\end{IEEEbiography}
\vspace{-20mm}
\begin{IEEEbiography}[{\includegraphics[width=.9in,clip,keepaspectratio]{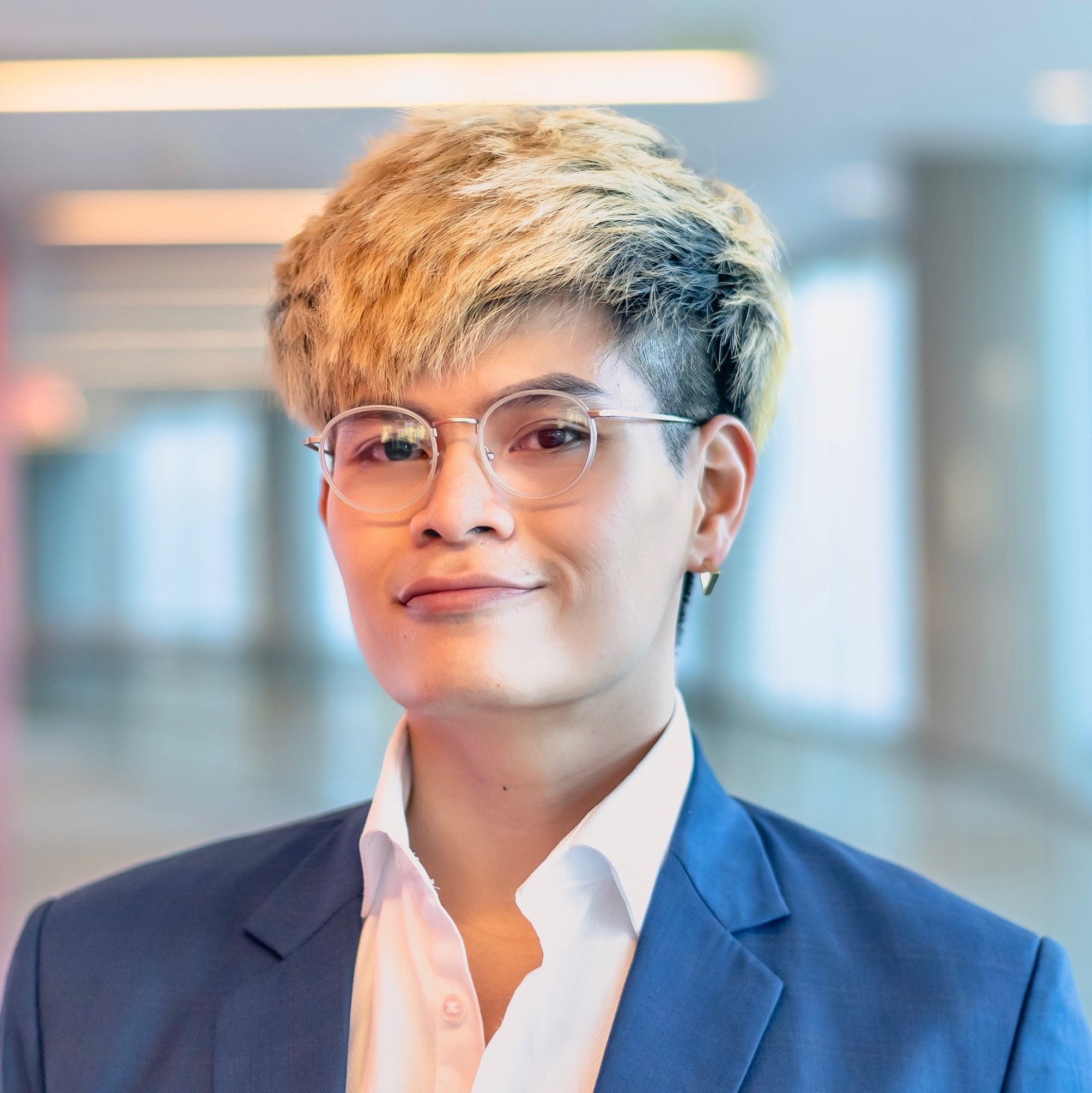}}]{Huy Tu} holds a Ph.D. in Computer Science from North
Carolina State University, Raleigh, NC. They explored frugal labeling 
processes while improving the data quality for software analytics. Now, they works for Meta Platforms, Inc.  \newline \url{https://kentu.us}.
\end{IEEEbiography}
\vspace{-20mm}
\begin{IEEEbiography}[{\includegraphics[width=.9in,clip,keepaspectratio]{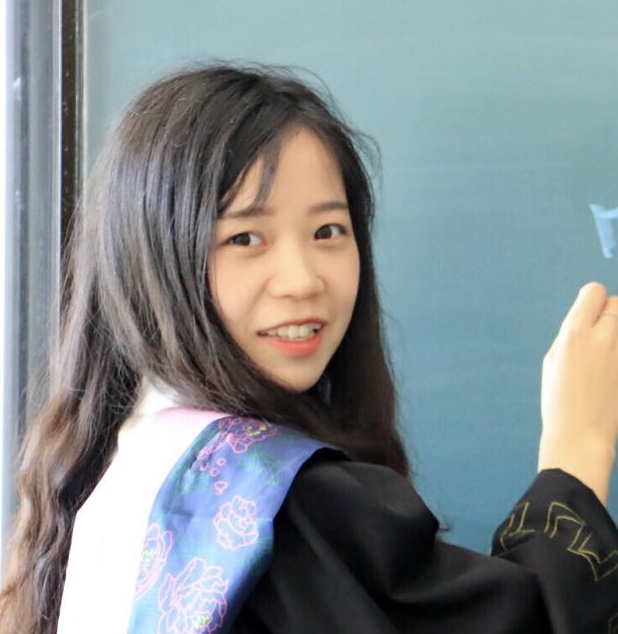}}]{Xueqi Yang} is a Ph.D. student in Computer Science at North Carolina State University.  Her research interests include automatic static analysis and applying human-assisted AI algorithms in software engineering.  \url{https://xueqiyang.github.io/}.
\end{IEEEbiography}
\vspace{-20mm}
\begin{IEEEbiography}[{\includegraphics[width=.8in,clip,keepaspectratio]{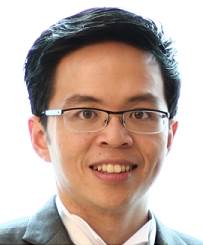}}]{David Lo} is a Professor in Computer Science at Singapore Management University. His research interests include software analytics, empirical software engineering, cybersecurity, and SE4AI.   \url{http://www.mysmu.edu/faculty/davidlo/}.
\end{IEEEbiography}
\vspace{-20mm}
\begin{IEEEbiography}[{\includegraphics[width=.9in,clip,keepaspectratio]{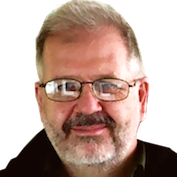}}]{Tim Menzies} (IEEE Fellow, Ph.D. UNSW, 1995)
is a Professor in Computer Science  at NC State University, USA.
His research interests include software engineering (SE), data mining, artificial intelligence, and search-based SE, open access science.  \url{http://menzies.us}.
\end{IEEEbiography}

\end{document}